\documentclass[revtex4,pre, preprint, epsfig, showpacs, superscriptaddress]{revtex4}

\usepackage[dvips]{graphicx}
\usepackage{ifpdf}
\usepackage{latexsym}
\usepackage{natbib}
\usepackage[dvips,
bookmarks,
bookmarksopen,
backref,
colorlinks,linkcolor={blue},citecolor={blue},urlcolor={blue},
]{hyperref}
\usepackage{subfig}

\begin{document}
\bibliographystyle{apsrev}

\title{Structure of Business Firm Networks and Scale-Free Models}

\author{Maksim Kitsak}
\affiliation{Center for Polymer Studies, Boston University,
Boston, Massachusetts 02215, USA}

\author {Massimo Riccaboni}
\affiliation{CUSAS, University of Florence, Florence, Italy}

\author{Shlomo Havlin}
\affiliation{Center for Polymer Studies, Boston University,
Boston,
  Massachusetts 02215, USA}
\affiliation{Minerva Center and Department of Physics, Bar-Ilan
  University, Ramat Gan, Israel}

\author{Fabio Pammolli}
\affiliation{Center for Polymer Studies, Boston University,
Boston, Massachusetts 02215, USA}

\affiliation{CUSAS, University of Florence, Florence, Italy}

\affiliation{IMT Institute for Advanced Studies, Lucca, Italy}

\author{H. Eugene Stanley}
\affiliation{Center for Polymer Studies, Boston University,
Boston, Massachusetts 02215, USA}


\begin{abstract}
We study the structure of business firm networks and scale-free models  with degree distribution $P(q) \propto
(q+c)^{-\lambda}$ using the method of $k$-shell decomposition. We find that the Life Sciences industry network consists of three components: a ``nucleus,'' which is a small well connected subgraph, ``tendrils,'' which are small
subgraphs consisting of small degree nodes connected exclusively to the
nucleus, and a ``bulk body'' which consists of the majority of nodes.
At the same time we do not observe the above structure in the Information and Communication Technology sector of industry.
We also conduct a systematic study of these three components in random
scale-free networks. Our results suggest that the sizes of the nucleus
and the tendrils decrease as $\lambda$ increases and disappear for
$\lambda \geq 3$. We compare the $k$-shell structure of random
scale-free model networks with two real world business firm networks in
the Life Sciences and in the Information and Communication Technology
sectors. Our results suggest that the observed behavior of the $k$-shell
structure in the two industries is consistent with a recently proposed
growth model that assumes the coexistence of both preferential and
random agreements in the evolution of industrial networks.

\end{abstract}



\maketitle

\section{Introduction}
Many real-world complex systems are often described using the
representation of graphs or networks, as sets of nodes connected by
links \cite{ER,ER_random_graphs,bollobas,wasserman, evolution,
  structure, handbook}. Networks appear in various areas of science such
as physics, biology, computer sciences, economics and sociology
\cite{bollobas,wasserman,evolution,structure,handbook}. Real networks,
despite their diversity, appear to have many common properties. Some
real networks have been found to be ``small-world''
\cite{collect_dynam,sw_problem,bollobas,diam_of_www}: despite their
large size, the typical distance between nodes is very small, of order
of $\log(N)$ or less \cite{bollobas,scaling1,compl_nw}, where $N$ is the
number of nodes. Also, some real networks are scale-free (SF) with a
power-law tail in their degree distribution
\begin{equation}
P(q) \propto(q+c)^{-\lambda},
\end{equation}
where $q$ is the number of links per node, $\lambda$ and $c$ are distribution parameters \cite{sf1, sf2, sf3, Barabasi_rmp_review}.

Many statistical physics and mathematical techniques have been
successfully applied to study network structure including percolation
\cite{compl_nw, percolation1, percolation2, percolation3, percolation4,
  robustness}, scaling \cite{structure, handbook, scaling1},
partitioning \cite{partitioning}, box covering \cite{self-sim, repulsion,
  bc, skeleton_and_fractal}, $k$-core percolation \cite{kcore1, kcore2}
and $k$-shell decomposition \cite{kshell} . The latter has been recently used to study the topology of the
Internet at the autonomous system (AS) level \cite{medusa}. It has been
found that the AS Internet network consists of three distinct
components:

\begin{itemize}

\item[{(i)}] The nucleus consists of highly connected nodes (hubs) and
  constitutes a tiny fraction of the network ($\approx 0.5 \% $).

\item[{(ii)}] The tendrils ($\approx 25 \% $) are the subgraphs that
  connect to the network exclusively via the nucleus.

\item[{(iii)}] The remaining nodes ($\approx 75\%$) of the network
  constitute the bulk body.  Unlike the tendrils, the nodes in the bulk
  remain connected even if the nucleus is removed.

\end{itemize}

\noindent The appearance of the three components led to the association
of the AS Internet structure with that of a jelly-fish model
\cite{medusa}. For an alternative jelly-fish structure see also
Ref.~\cite{jellyfish}. The nucleus plays a crucial role in data transfer
throughout the Internet. If the nucleus is damaged or blocked, nodes in
the tendrils cease to communicate with other nodes both in the bulk body
and in other tendrils. However, the nodes in the bulk body can still
communicate with each other, but less efficiently since the average path
length between them almost doubles \cite{medusa}. The network of
workplaces in Sweden \cite{fragmentation, sweden} was also shown to
possess a jelly-fish structure. Mean field analysis of the $k$-shell
structure \cite{kcore, bootstrap} focuses on random networks with
a given degree distribution $P(q)$.

In the present work we address the long-standing question of how long does a typical leader in an industry maintain its position. This
question has attracted continuing attention in the Industrial Organization literature over the past generation \cite{sutton}.
Two rival views exist. The first asserts that leadership tends to persist for a 'long' time while the second one emphasizes the transience
of leadership positions due to radical innovation and "leapfrogging competition". The Life Sciences (LS) sector is generally cited as an archetypal example of stability of industry leaders while on the contrary the Information and Communication Technology (ICT) sector is widely considered as a good example of high market turnover and instability. We study the $k$-shell structure of the LS and ICT industry in order to test these views and better understand the reasons behind the higher stability of the leading firms in the LS sciences and the ICT sector. The networks of LS and ICT firms are networks of nodes representing firms in the worldwide LS and ICT industries with links representing the collaborative agreements among them \cite{bc, pharma, pharm1, ict}. We also conduct a numerical analysis of model SF networks in order to better understand their $k$-shell structure. We study how the size and the connectivity of the three components of random SF model networks depend on the SF parameters. We compare the $k$-shell structure of SF models with that of the LS and ICT industry networks in order to get a better insight into the growth principles of the Industry networks.

The rest of the manuscript is organized as follows: in
Section II we define the k-shell decomposition and apply it to
both real and model SF networks. We then analyze the leadership in the industry networks from the $k$-shell perspective. In Section III we conduct a
systematic analysis of the properties of the three components of
model SF networks. In section IV we compare the results for SF models
with those observed in the two industrial networks.
We conclude our manuscript in Section V with a discussion and
summary.

\section{$K$-shell, $K$-core, and $K$-crust}

We start the process of the $k$-shell decomposition \cite{medusa, bootstrap} on a network by removing all nodes
with degree $q=1$. After the first iteration of pruning, there may
appear new nodes with degrees $q=1$. We keep on pruning these nodes
until only nodes with degree $q\geq2$ are left. The removed
nodes along with the links connecting them form the $k=1$ shell.  Next,
we iterate the pruning process for nodes of degree $q=2$, thereby
creating the $k=2$ shell. We further continue the $k$-shell decomposition for higher values of $q$ until all nodes of the network are
removed. As a result each node in the network is assigned a $k$-shell index
$k$. The largest shell index is called $k_{\rm max}$, which is also the
total number of shells in the network, provided all shells below $k_{\rm
  max}$ exist. The $k$-crust is defined as the union of all $k$-shells with
indices smaller than $k$. Similarly, the $k$-core is defined as the
union of all nodes with indices greater or equal $k$. (See Fig. \ref{drawing}
for demonstration.) As we explain later, the set of nodes in the $k_{\rm max}$
shell is called nucleus provided there is a large number of nodes
(tendrils) connected to the network exclusively via the $k_{\rm max}$ shell.

\begin{figure} [!ht]
\includegraphics[width=8.0 cm,height=7.0cm,angle=0]{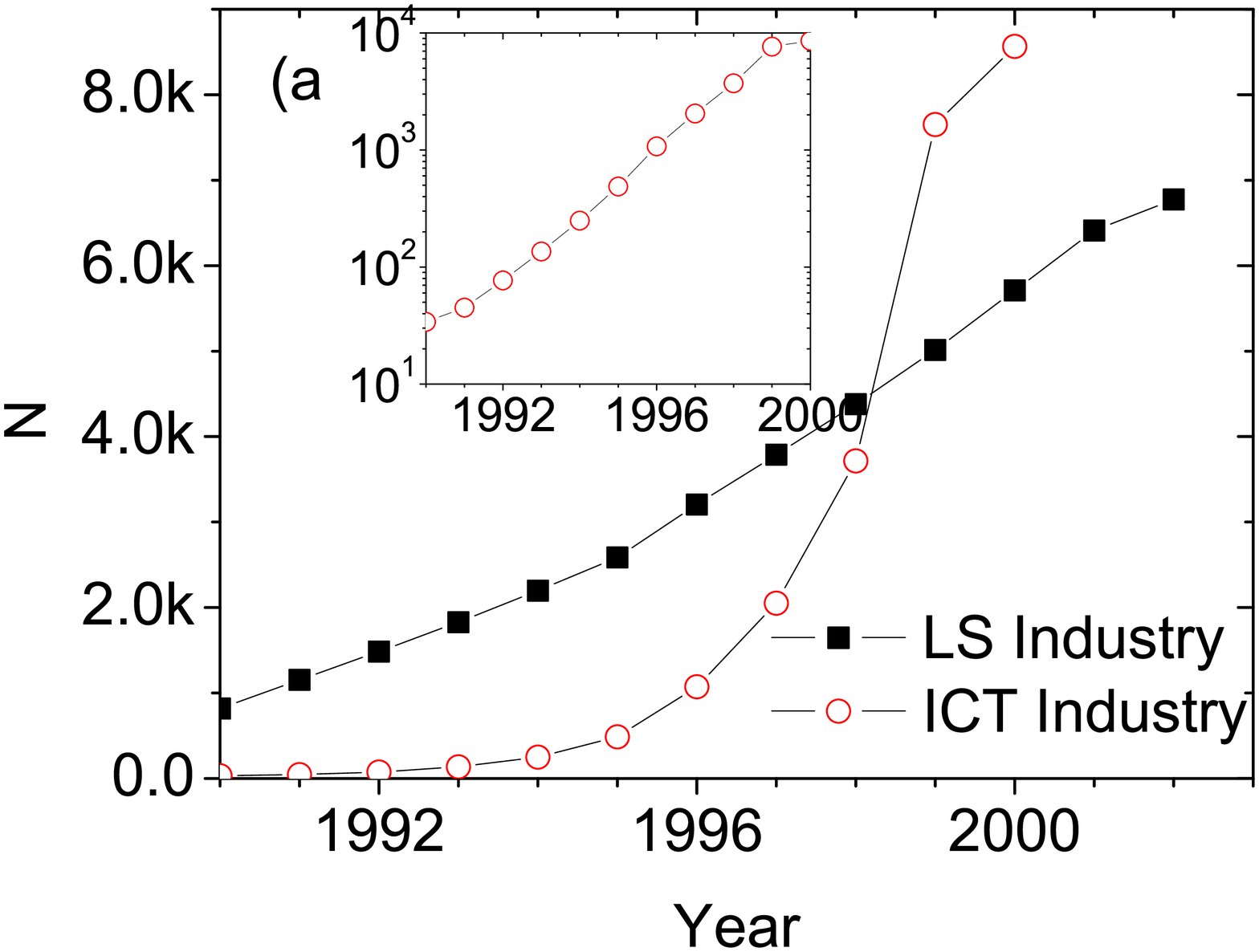}
\includegraphics[width=8.0 cm,height=7.0cm,angle=0]{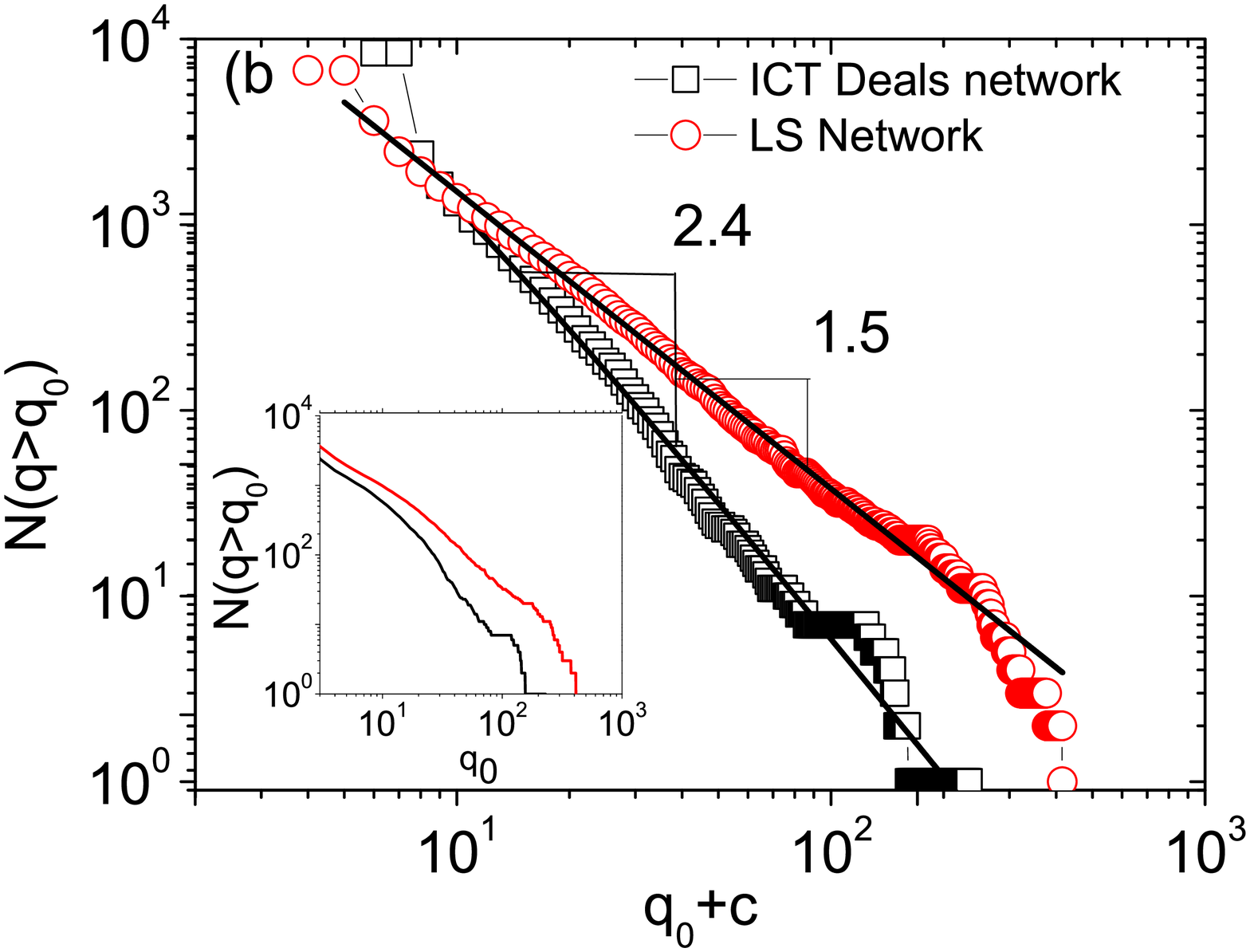}
\caption{(a) The growth of the largest connected component of LS and ICT
  industries. The LS industry expands almost linearly while the ICT
  industry exhibits exponential growth (see inset, a semi-log plot of
  the ICT network size as a function of time).  (b) Cumulative degree
  distribution $N(q>q_{0})$ of LS and ICT
  networks. The inset displays cumulative degree distribution $N(q>q_{0})$ of LS and ICT network without offset $c$.}
\label{growth}
\end{figure}

We analyze the $k$-shell structure of LS and ICT industrial sectors in
time periods between 1990 and 2002 and between 1990 and 2000
respectively. The LS network expanded linearly since the mid-1970s while
the ICT network took off in the 1990s and grew exponentially for a
decade [see Fig.~\ref{growth}(a) and the inset].  The total number of
firms in the LS is $N=6,776$ and in the ICT is $N=7,759$. These
sizes refer to the largest connected component of each of the networks
at the last year of observation.  Both industrial networks feature SF
degree distributions with $\lambda \approx 2.5$, $c \approx 4.$ (LS) and
$\lambda \approx 3.4$, $c \approx 6.$ (ICT) [see
  Fig.~\ref{growth}(b)]. We use the Kholmogorov-Smirnov test in order to examine the goodness of fit of the degree distributions. The obtained p-values for $1000$ trials are $0.24$ and $0.33$ for the LS and the ICT industry networks respectively.

\begin{figure}[!h]
\includegraphics[width=8.0 cm,height=7.0cm,angle=0]{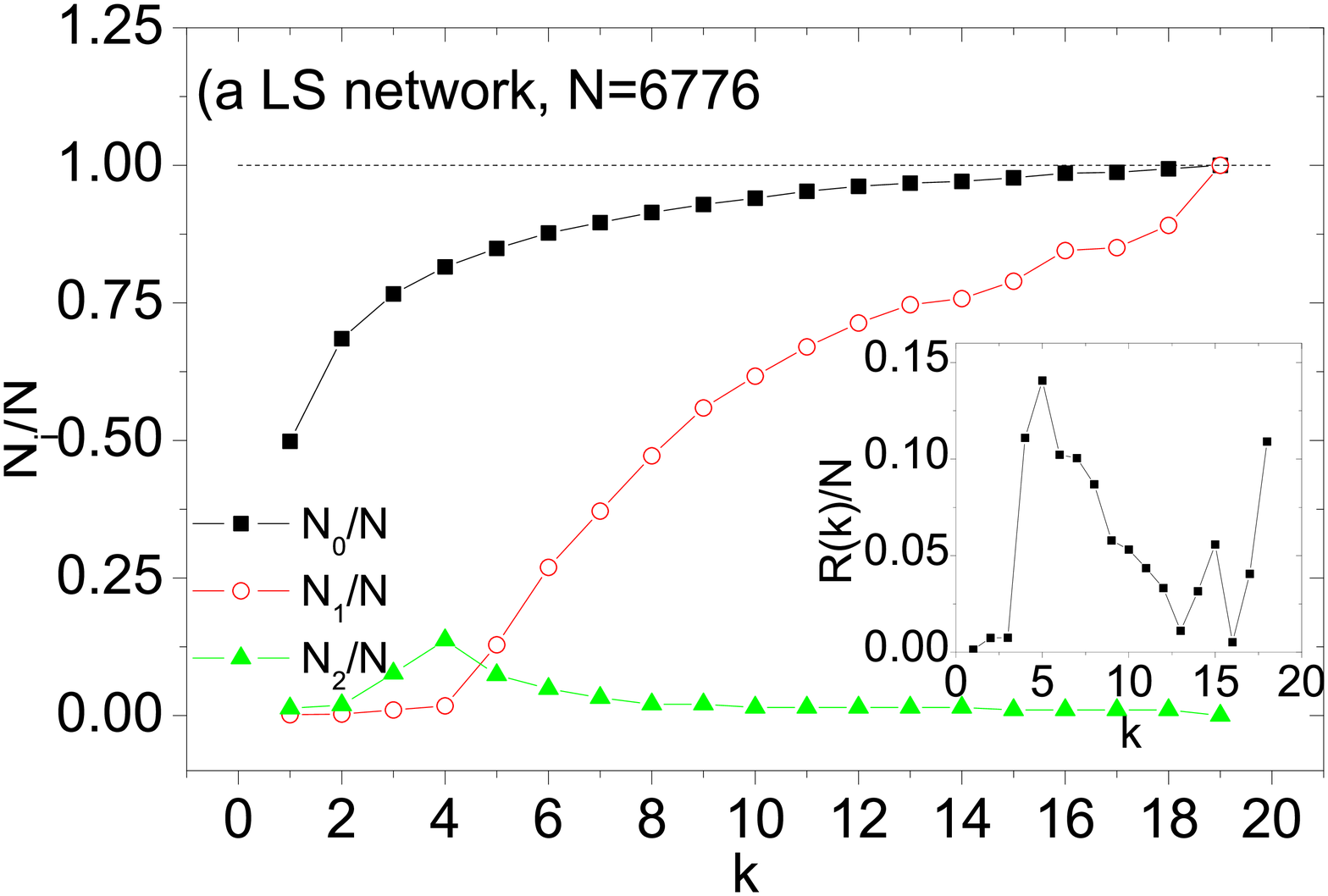}
\includegraphics[width=8.0 cm,height=7.0cm,angle=0]{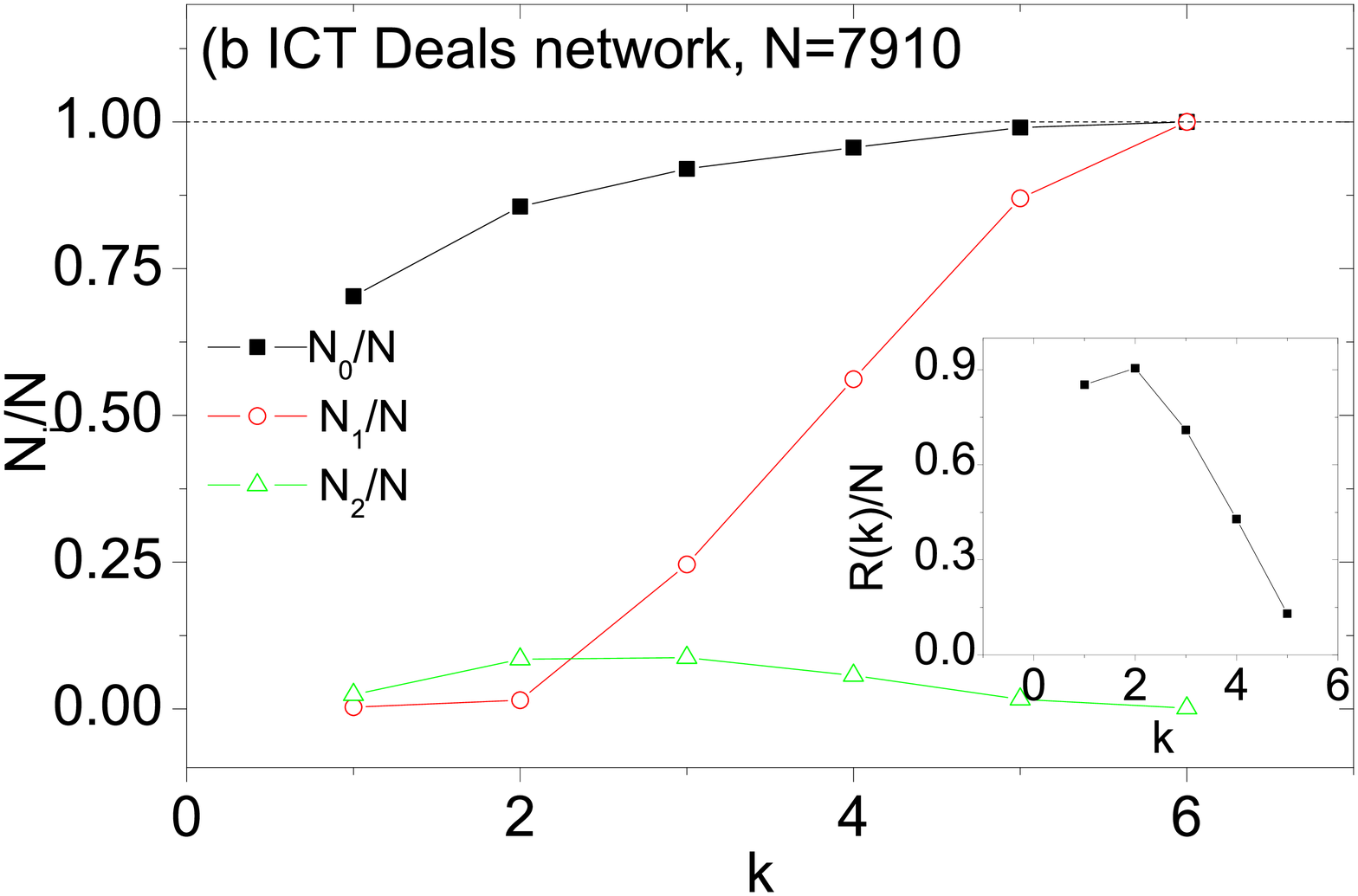}
\includegraphics[width=8.0 cm,height=7.0cm,angle=0]{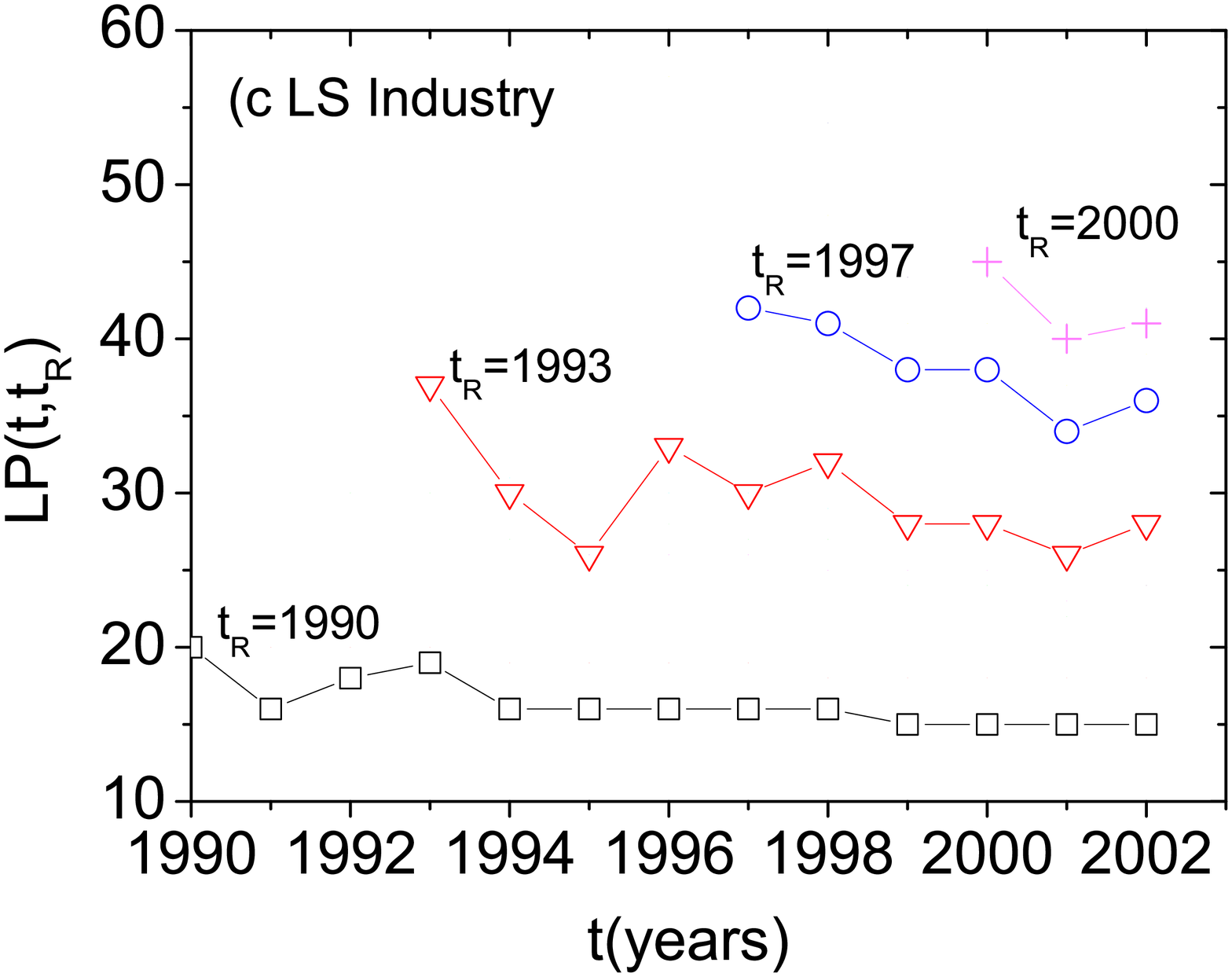}
\caption{(a) The $k$-shell structure of the LS network and (b) of the
  ICT network. Plots of the size of the $k$-crust, $N_{0}$; the size of
  the largest connected component of the $k$-crust, $N_{1}$; and the
  size of the second largest connected component of the $k$-crust,
  $N_{2}$, as a function of the $k$-crust index $k$. The sizes are
  normalized with the total number of nodes in the network, $N$. $N_{2}$
  is multiplied by 100. The transition of the $k$-crust at shell $k=18$
  of the LS network reveals a jelly-fish topology \cite{medusa}. Since
  the change in $F_{1} \equiv N_{1}/N$ at the $k=k_{\rm max}$ is smaller
  in the ICT network than in the previous shell $k=k_{\rm max} - 1$, the ICT does not have a jelly-fish structure. The
  insets of (a,b) show the rate $R$ of the largest connected component
  change as a function of shell index $k$. (c) The leadership persistence $LP(t,t_{R})$ as a function of years in the LS industry.  We define $LP(t,t_{R})$ as a number of firms that were in the nucleus both at time $t_{R}$ and $t$. Note that most of the LS firms remain in the nucleus after they first enter it.}
\label{fig_real_kshells}
\end{figure}

We next apply the $k$-shell decomposition procedure to the LS and the
ICT networks.  For each $k$-crust we calculate the total number of
nodes, $N_{0}$, comprising it, the size of the largest connected
component, $N_{1}$, in the $k$-crust and the size of the second largest
component, $N_{2}$, of the $k$-crust.  As seen in Fig. \ref{fig_real_kshells},
the LS network consists of $k_{\rm max} =19$ shells while the ICT
network (which consists of comparable number of firms) has only $k_{\rm
  max} =6$ shells. The size of the largest cluster in the $k$-crust
starts growing rapidly after $k=4$ and $k=2$ in the LS and the ICT
networks respectively. At these values of $k$, the size of the second
largest cluster in the $k$-crust reaches a maximum. The above behavior of the $k$-crust components is consistent with the
existence of  a second order phase transition in the $k$-crust
structure \cite{medusa}. The type of the phase transition taking place at the
$k$-shell decomposition is similar to that of targeted percolation in SF
networks \cite{robustness, breakdown}.

Unlike in the ICT network,
the size of the largest connected component $N_{1}$ in LS network
undergoes a large jump $N_{1}(k=19)-N_{1}(k=18)= 745$, while the total size $N_{0}$ of the crust
experiences only a small change $N_{0}(k=19)-N_{0}(k=18)= 43$ (See Fig. \ref{fig_real_kshells}(a)). This can be explained as follows \cite{medusa}.
Approximately $10\% $ of the he LS network are firms (which we call "tendrils") that prefer to sign collaborative agreements exclusively
with the $43$ firms that form $k=k_{max}=19$ shell (which we call nucleus). Typically, each of these firms (tendrils) signs a small number of agreements with firms in the nucleus and, therefore, has small degree.  Thus, the tendrils are removed in the decomposition of the first few
shells. However, being connected exclusively to the nucleus in  the $k_{\rm max}=19$ shell,
the tendrils do not contribute to the largest connected component of the
$k$-crust until the last $k_{\rm max}$ shell is decomposed. It is the inclusion of the tendrils firms into $N_{1}(k)$ at the decomposition of the $k_{max}$ shell that results in the observed jump in $N_{1}(k)$. The appearance of the three components---the nucleus, the tendrils and the bulk body---allows one to associate the structure of the network of LS firms with that of the jelly-fish similar to the Internet at the Autonomous System level \cite{medusa}. Interestingly, we do not observe the a similar jump in $N_{1}(k)$ in the ICT network (See Fig. \ref{fig_real_kshells}(b)).

In both LS and ICT industry sectors the $k_{max}$ shells include market leaders such as
Pfizer, GSK, Novartis, J\&J, Sanofi-Aventis, Bayer in the LS network and
Microsoft, IBM, AT\&T, Yahoo, Cisco, AOL, Time Warner and Google in the ICT
network.  We find that the LS network tendrils are mostly composed of the new start up firms
and their university partners. On the one hand, start up firms preferentially attach to market leaders. On the other hand, market leaders
compete to sign exclusive deals with new and promising start up
firms. We also notice the remarkable stability of the LS nucleus: once a particular firm enters the nucleus it is very likely to remain there for many years.(See Fig. \ref{fig_real_kshells}c) On the contrary, in the ICT sector there is more emphasis on how
to integrate different technologies and markets \cite{ict}. Thus, there is more deals between firms of the same size and age but in different technological
and market areas. As a consequence, the $k$-shell of the ICT network is more unstable and heterogeneous which may explain why we do not detect the emergence of a nucleus such as in the LS sector.

As discussed above, one can in general calculate the size of the nucleus $S_{n}$ as the total number of nodes in the $k_{max}$ shell

\begin{equation}
S_{n}=N_{0}(k_{\rm max})-N_{0}(k_{\rm max}-1).
\end{equation}
The increase in $N_1(k)$  at $k=k_{max}$ is comprised by the inclusion of tendrils $S_{t}$ and the nucleus $S_{n}$. Thus, the size of tendrils can be calculated as
\begin{equation}
S_{t}=N_{1}(k_{\rm max})-N_{1}(k_{\rm max}-1)-S_{n}.
\end{equation}

\begin{figure}[!h]
\includegraphics[width=8.0 cm,height=7.0cm,angle=0]{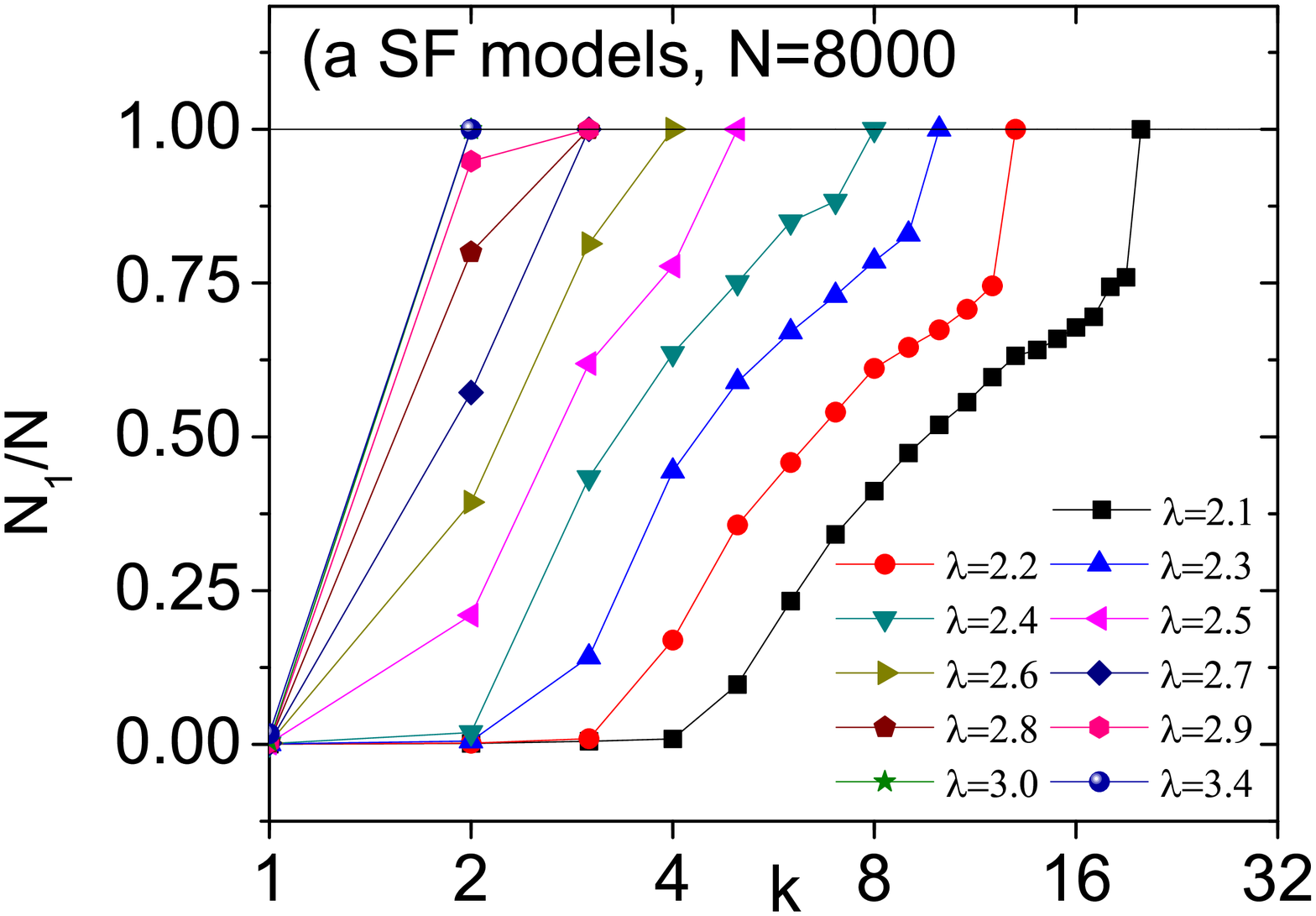}
\includegraphics[width=8.0 cm,height=7.0cm,angle=0]{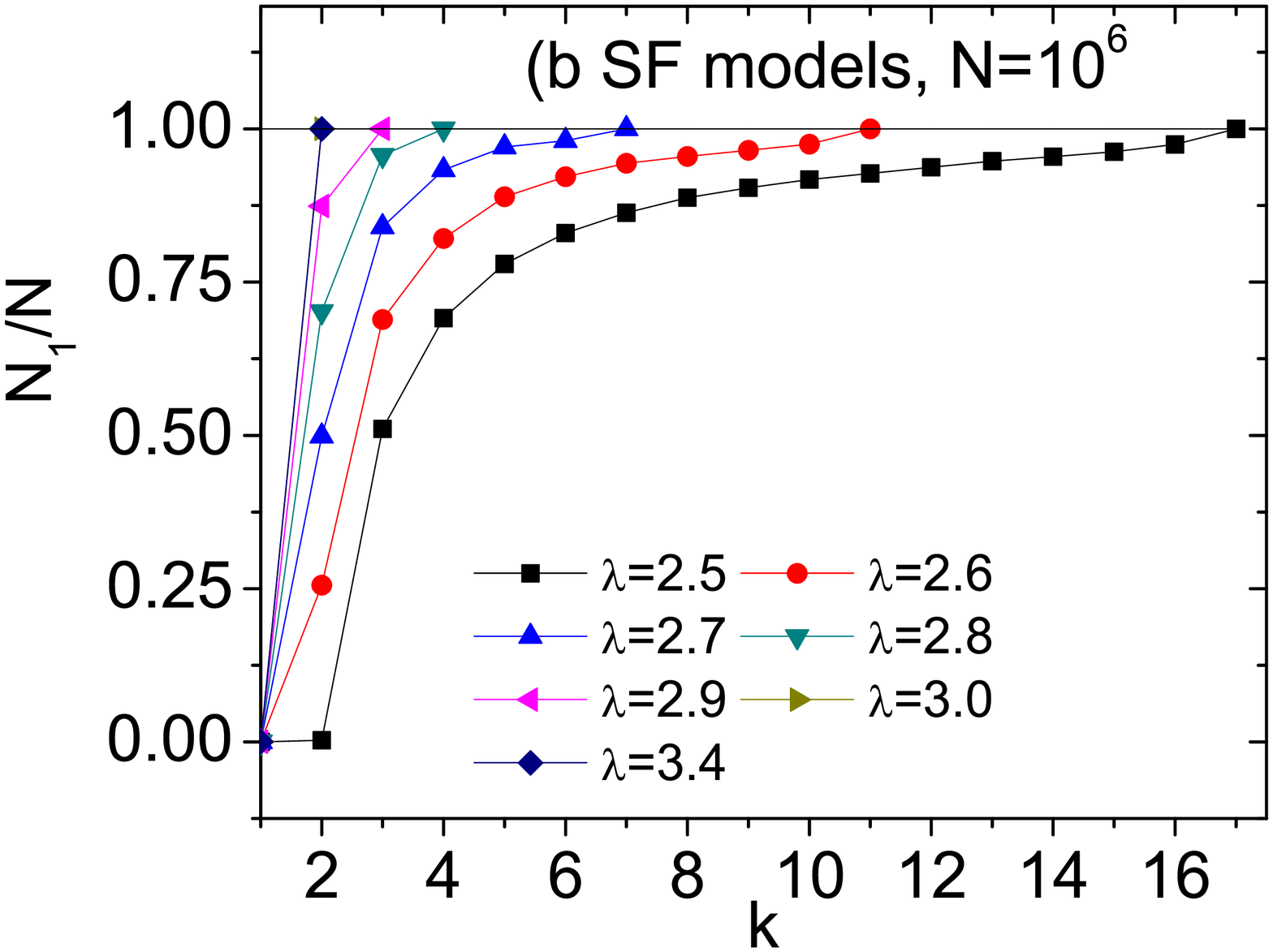}
\includegraphics[width=8.0 cm,height=7.0cm,angle=0]{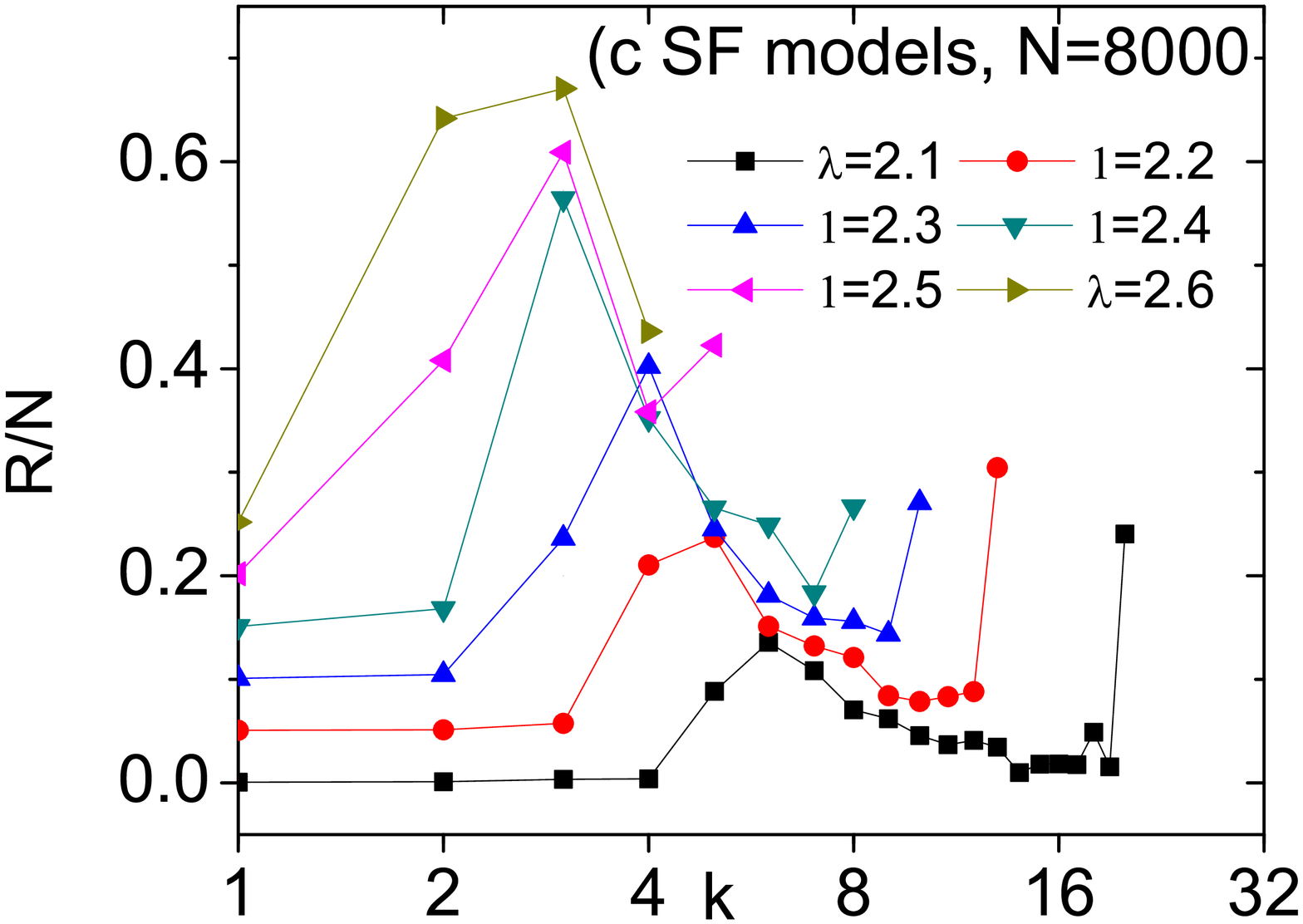}
\includegraphics[width=8.0 cm,height=7.0cm,angle=0]{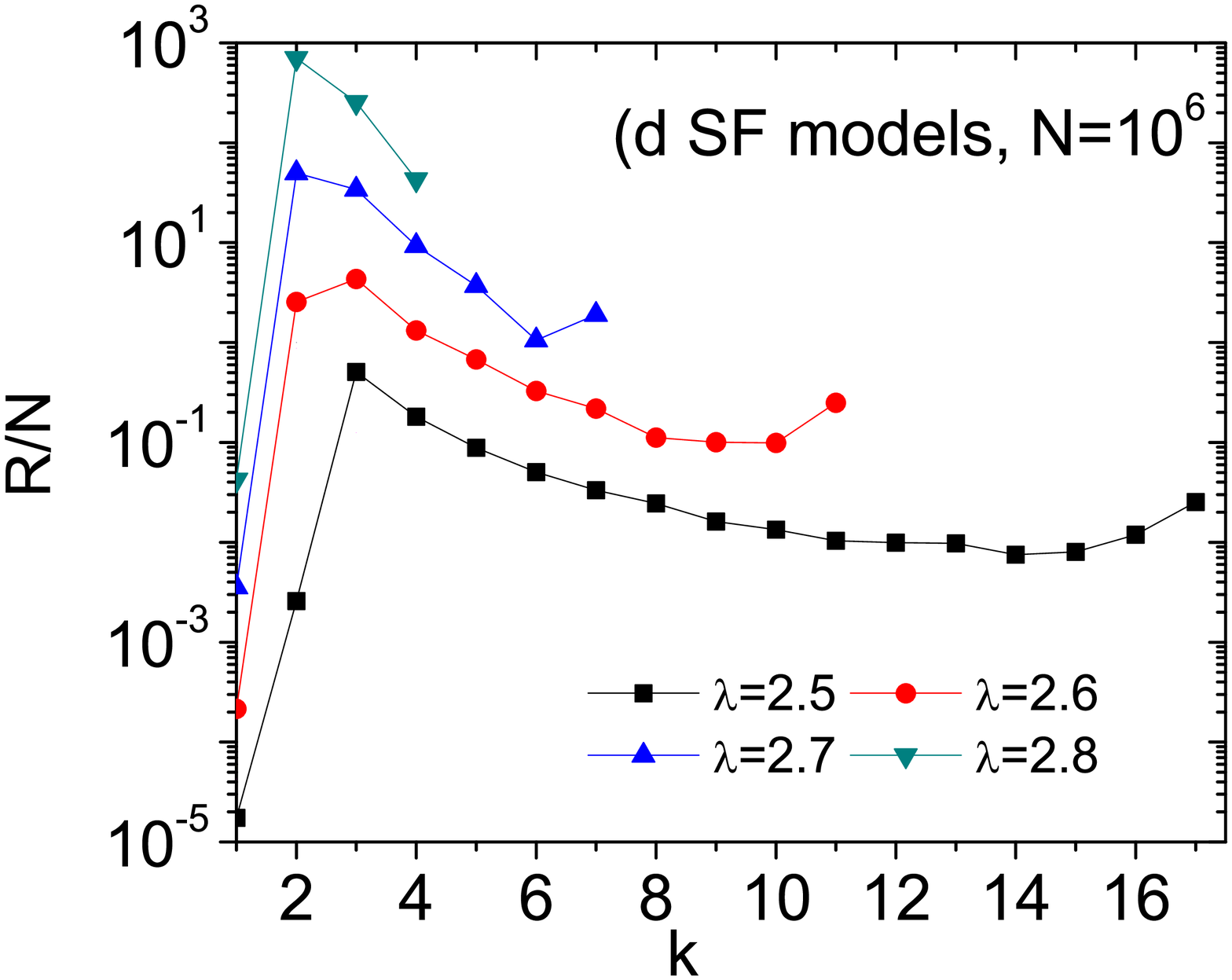}
\caption{(a,b) The $k$-shell structure
  of random SF models with (a) $N=8000$ and (b) $N=10^{6}$ nodes are
  shown for comparison.  Note that $\lambda = 3.4$ curve overlaps with that of $\lambda = 3.0$. (c,d) The rate of the largest connected component change, $R$, as a function of shell index $k$ for (c) $N=8000$ and (d) $N=10^{6}$. Note that in order to avoid the overlap of curves in (c,d) we subsequently shift the plots with respect to each other by the additive factor $0.05$ in (c) and the multiplicative factor of $10$ in (d).
  The sizes of the  nucleus and the tendrils decrease as $\lambda$ increases. The nucleus
  and the tendrils disappear in (a,c) for $N=8000$ at $\lambda > 2.5$ and in
  (b,d) for $N=10^{6}$ at $\lambda > 2.7$.}
\label{fig_kshells}
\end{figure}

As seen in Fig. \ref{growth}(b), both the LS and the ICT industrial networks exhibit a SF degree
distribution. Hence, in order to better understand the substructures of
real networks---the nucleus, the tendrils, and the bulk components---we
analyze the $k$-shell structure of the random SF models which were
generated using the configurational approach \cite{sfmodel}. We
calculate the $k$-shell structure of random SF networks with $c \geq 0$, and degree distribution exponent $\lambda \in [2,3]$.
For our simulations (Fig. \ref{fig_kshells}) we choose networks sizes of $N=8,000$ (which is comparable to the size of the LS and ICT industry networks) and $N=10^{6}$ in order to test the influence of finite size effects.

It is seen in Fig. \ref{fig_kshells} that both the number of shells $k_{\rm max}$ and the jump in the largest connected component $N_{1}(k)$ decrease as the exponent $\lambda$ increases. As the jump becomes less pronounced it becomes harder to detect it which motivates us to introduce the quantitative criterion for the emergence of the three distinct components. We define the rate of change of the largest connected component size as
\begin{equation}
R(k)\equiv N_1(k)-N_1(k-1).
\end{equation}
We compare the increase of the largest connected component $R(k)$ at $k=k_{max}$ with that at $k=k_{max} - 1$. The jump in $N_{1}(k)$ results in
\begin{equation}
R(k_{\rm max})>R(k_{\rm max}-1).
\label{nucleus_def}
\end{equation}
We use Eq.~(\ref{nucleus_def}) as a criterion for the existence of a nucleus and tendrils.

By examining $R(k)$ plot we observe in the case of $N=8,000$ (Fig. \ref{fig_kshells}c), that SF
models with $\lambda > 2.5$ already do not have a nucleus and
tendrils. However, for $N=10^{6}$ we do observe nucleus and tendrils in SF models for
$\lambda \leq 2.7$.
(Fig. \ref{fig_kshells}d).  These observations suggest that all SF networks with $\lambda \in (2,3)$ have a nucleus
and tendrils provided $N$ is sufficiently large.
The above observations for SF models agree with the fact that we observe jelly-fish topology for
LS network ($\lambda=2.5$) and do not observe it for ICT network $\lambda=3.4$. However, SF model networks with $N=8000$ with $\lambda = 2.5$ and $\lambda = 3.4$
have $k_{max} = 5$ and $k_{max} = 2$ respectively, while the measured $k_{max}$ in LS and ICT networks (which have the same $\lambda$ values)
are $k_{max} = 19$ and $k_{max} = 6$. The observed difference in $k_{max}$ between industry networks and SF models with similar parameters suggests to further explore the $k$-shell structure of SF networks and consider SF model networks with $c>0$ values in $P(q)$ (as found in the LS and the ICT networks).

\section{$K$-Shell Properties of scale-free networks}

\begin{figure}[!h]
\includegraphics[width=7.0 cm,height=6.0cm,angle=0]{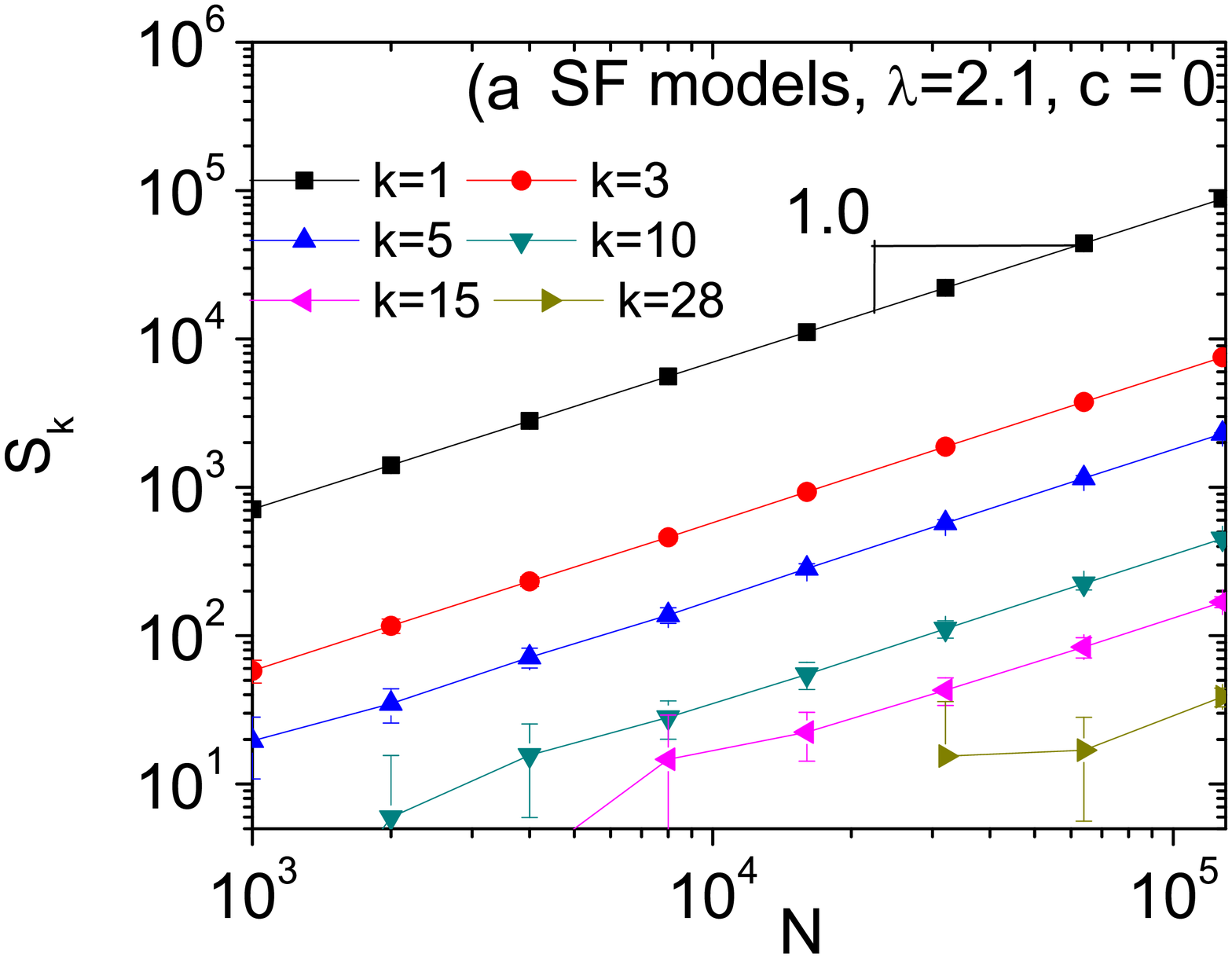}
\includegraphics[width=7.0 cm,height=6.0cm,angle=0]{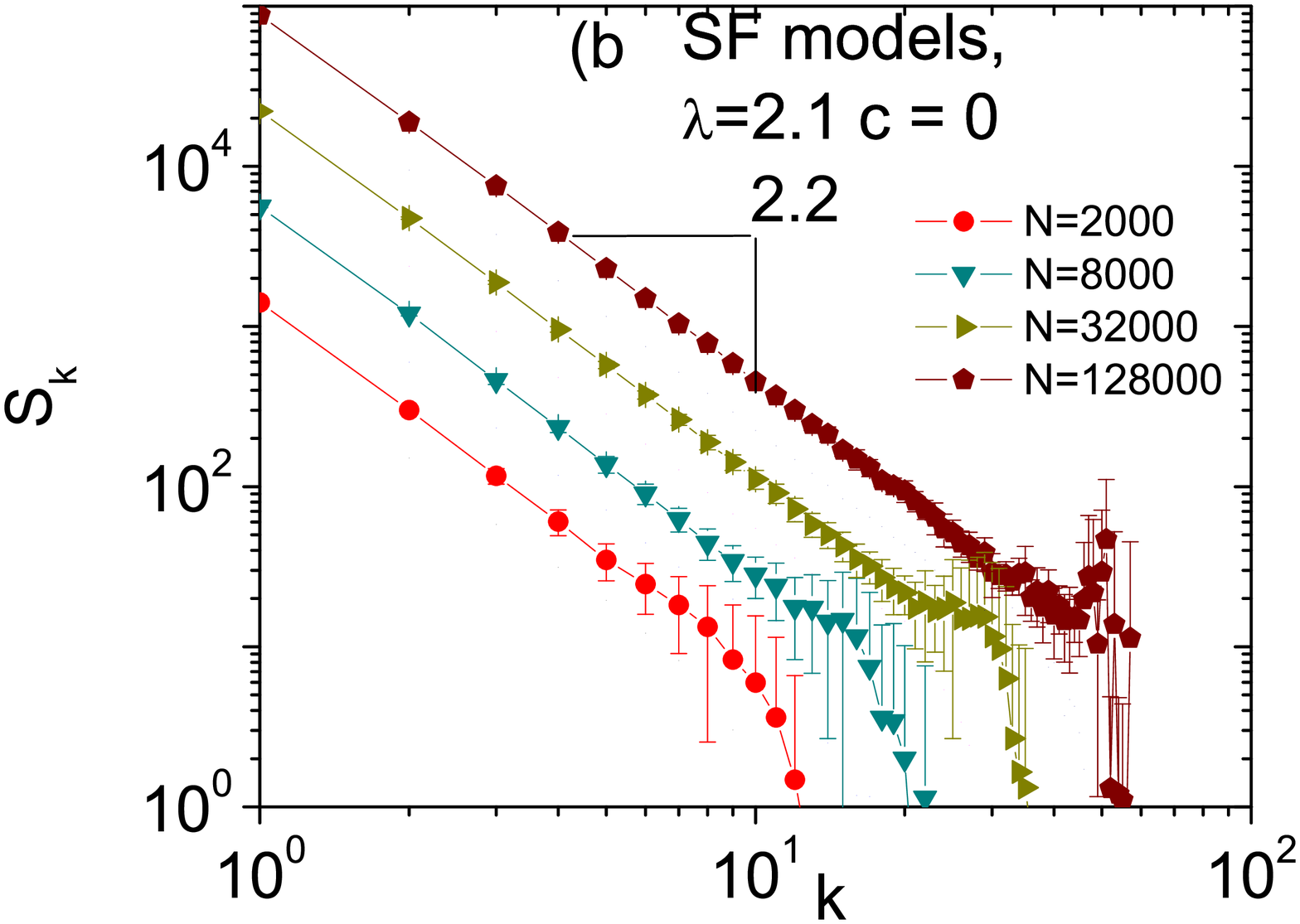}
\includegraphics[width=7.0 cm,height=6.0cm,angle=0]{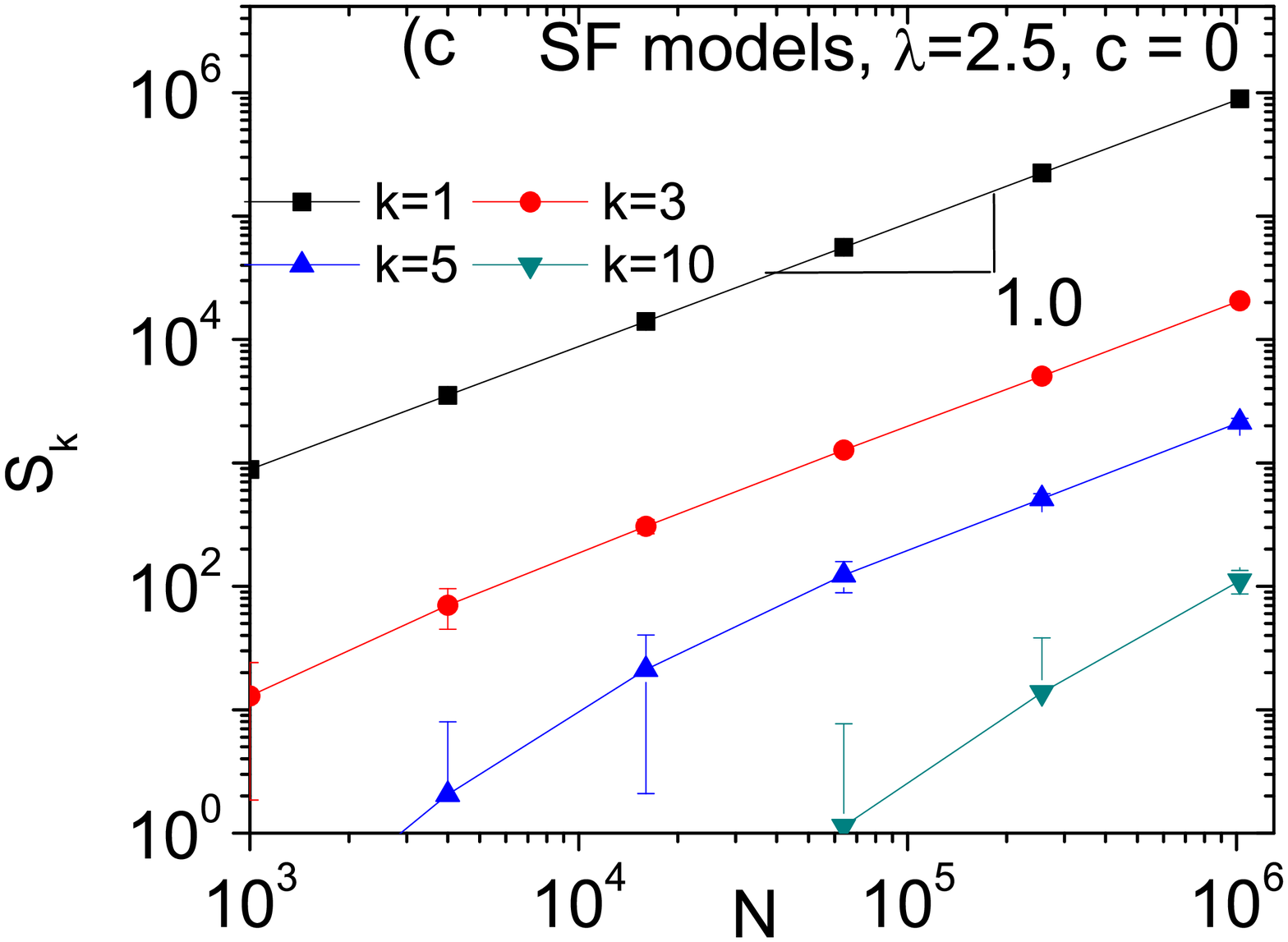}
\includegraphics[width=7.0 cm,height=6.0cm,angle=0]{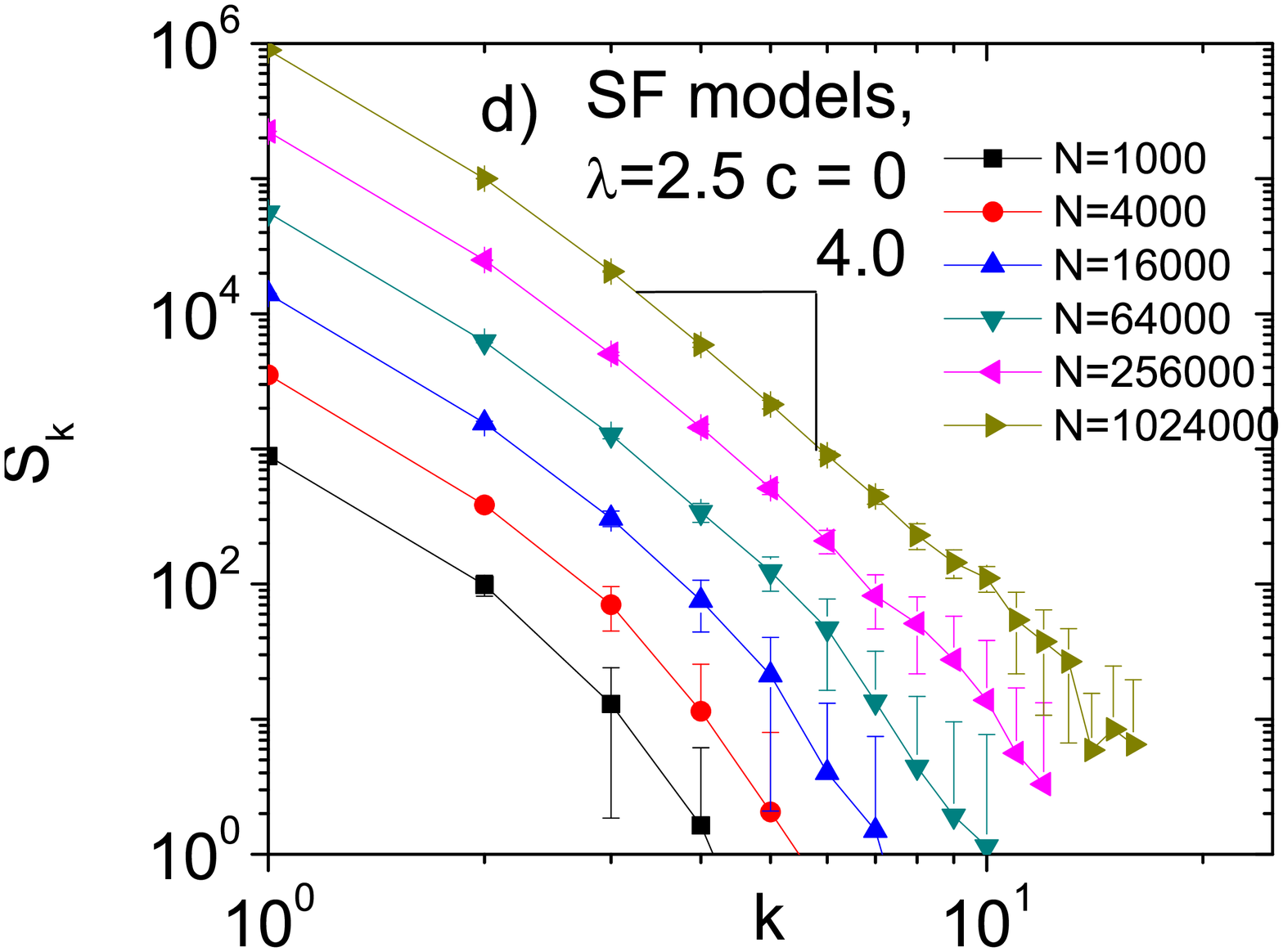}
\includegraphics[width=7.0 cm,height=6.0cm,angle=0]{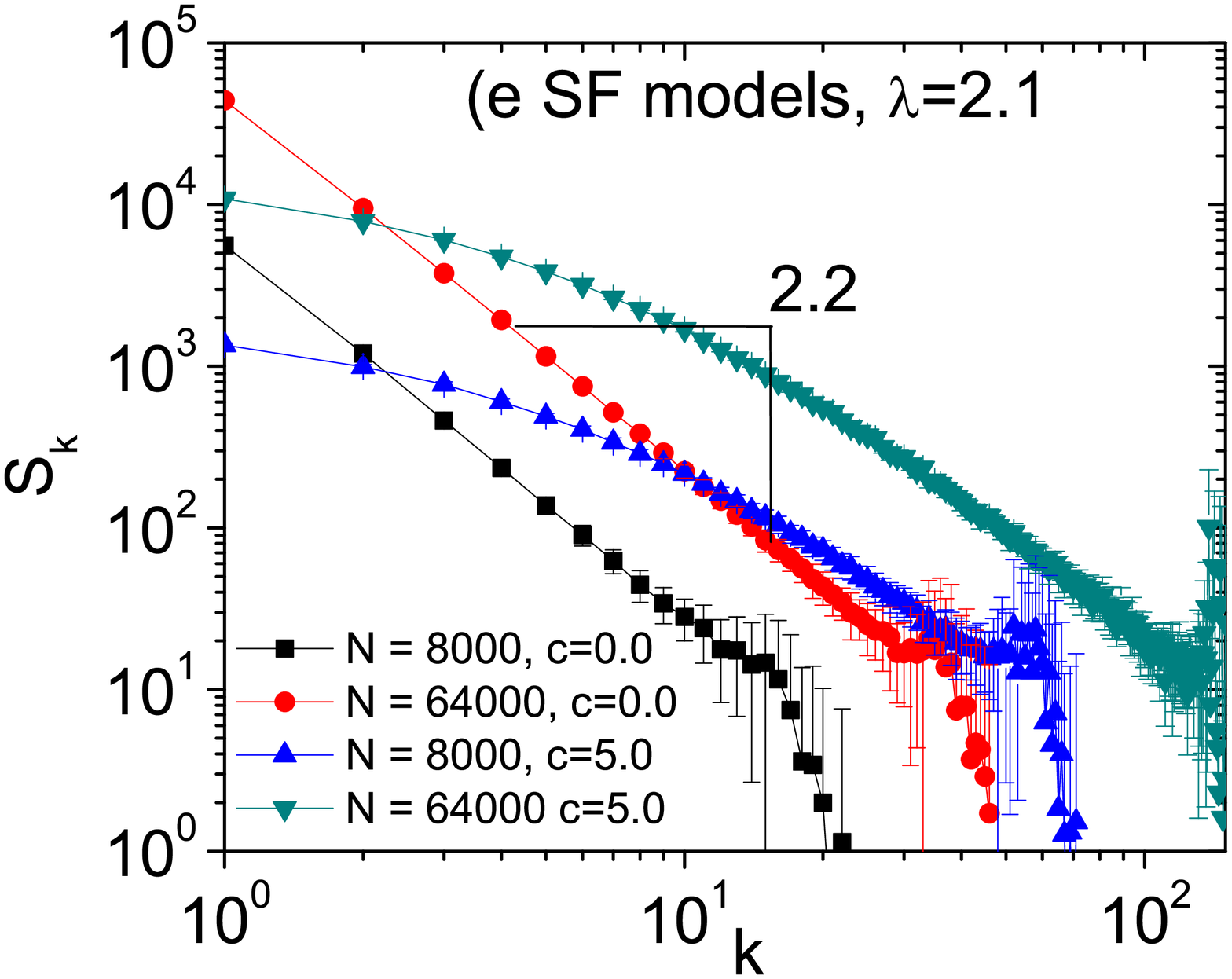}
\includegraphics[width=7.0 cm,height=6.0cm,angle=0]{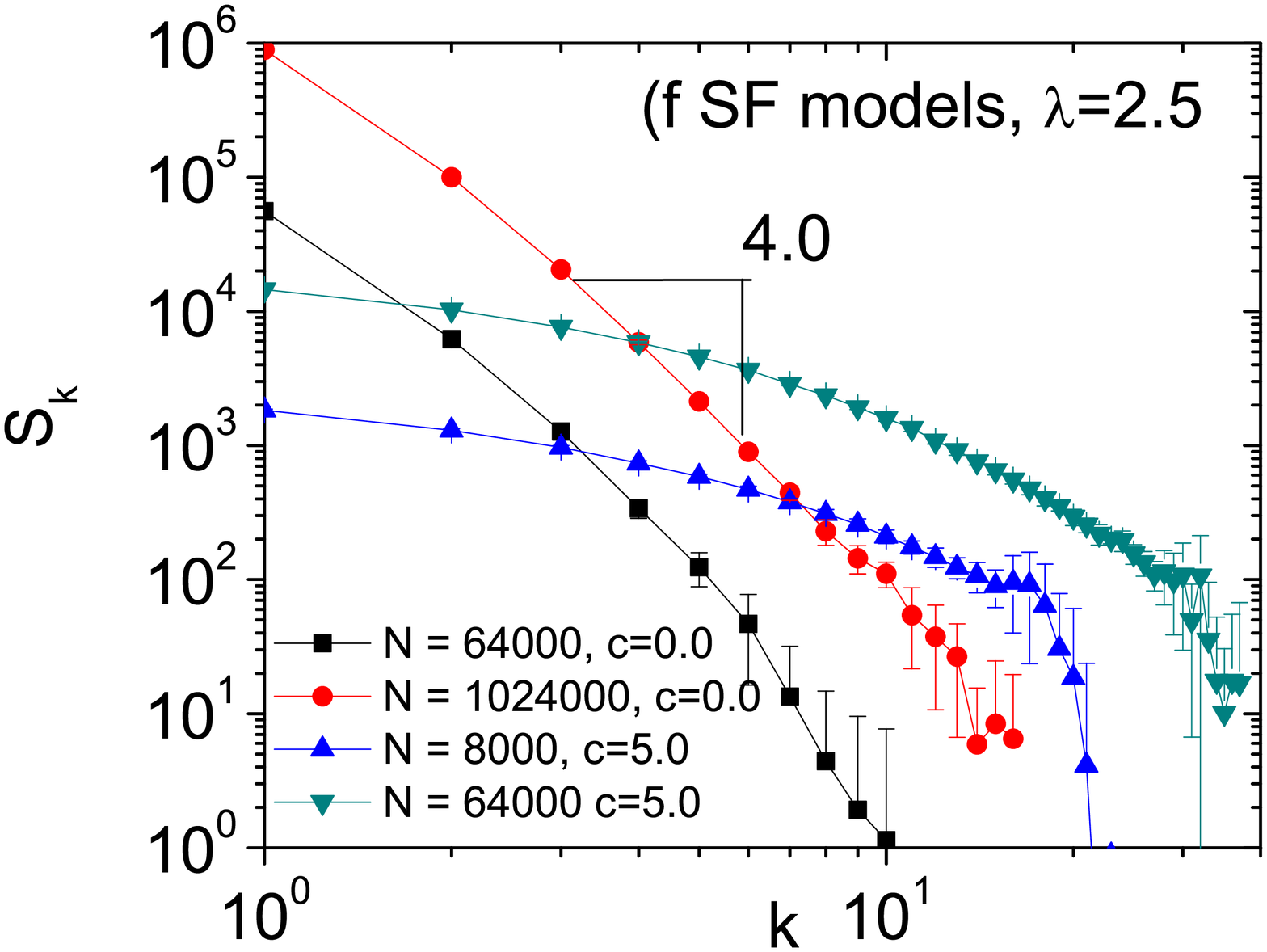}
\caption{Sizes of $k$-shells, $S_{k}$, for SF random models with $c=0$
  (a,b) $\lambda =2.1$ and (c,d) $\lambda = 2.5$ as a function of (a,c)
  $N$ and the $k$-shell index (b,d) $k$. Note that sizes of $k$-shells
  increases proportionally to $N$. (e,f) Sizes of $k$-shells, $S_{k}$,
  for SF networks with $c=5.0$ and (e) $\lambda =2.1$, (f) $\lambda =
  2.5$. SF models with $c=5.0$ have significantly larger number of
  shells, $k_{\rm max}$, compared to SF models with the same $\lambda$
  and $c=0$. $S_{k}$ in SF models with $c=5.0$ decreases significantly
  slower as a function of $k$ compared to SF models with the same
  $\lambda$ and $c=0$ in the small $k$ region. In the large $k$ region
  both types of SF models with $c=0.0$ and $c=5.0$ seem to have similar behavior
  $S_{k} \sim k^{-\delta}$, where $\delta = 2/ (3 - \lambda)$.}
\label{fig_sizes}
\end{figure}

As shown above, SF networks may or may not have a jelly-fish structure,
depending on the degree distribution $P(q)$ and size $N$.  In order to
better understand the $k$-shell structure of SF networks, we calculate
the sizes of the $k$-shells as a function of $N$ and $\lambda$. For each
pair of values $N$ and $\lambda$ we generate $10^{3}-10^{4}$
realizations of SF models and calculate the average number of shells
$k_{\rm max}$ constituting the network as well as their average size
$S_{k}$ (see Fig.~\ref{fig_sizes}). For small $k$ values, $S_{k}$ is
proportional to $N$. As the size of the network increases new shells
start to appear. When the size increases further, the new shell growth
stabilizes and becomes also proportional to $N$ (See
Fig.~\ref{fig_sizes}a,c). This result indicates that the size of each
shell constitutes a certain finite fraction of the network and this
fraction decreases with increasing $k$. The analytical analysis of $k$-core structure
\cite{bootstrap} leads to
\begin{equation}
S_k\propto k^{-\delta},
\end{equation}
where
\begin{equation}
\delta={2\over 3-\lambda}.
\end{equation}
Hence $\delta \approx 2.2$ for $\lambda = 2.1$ and $\delta \approx 4.0$
for $\lambda = 2.5$, which agrees with our simulations. [Figs.~\ref{fig_sizes}(b)
  and \ref{fig_sizes}(d)]. The appearance of $c>0$ in the SF degree
distribution significantly increases $k_{\rm max}$ [see
  Figs.~\ref{fig_sizes}(e) and \ref{fig_sizes}(f)]. However, the
asymptotic dependence of $k$-shell sizes seems to remain the same as $k$
approaches $k_{\rm max}$: $S_{k}\sim k^{-\delta}$
[Figs.~\ref{fig_sizes}(e) and \ref{fig_sizes}(f)].

\begin{figure}[!ht]
\includegraphics[width=8.0 cm,height=6.9cm,angle=0]{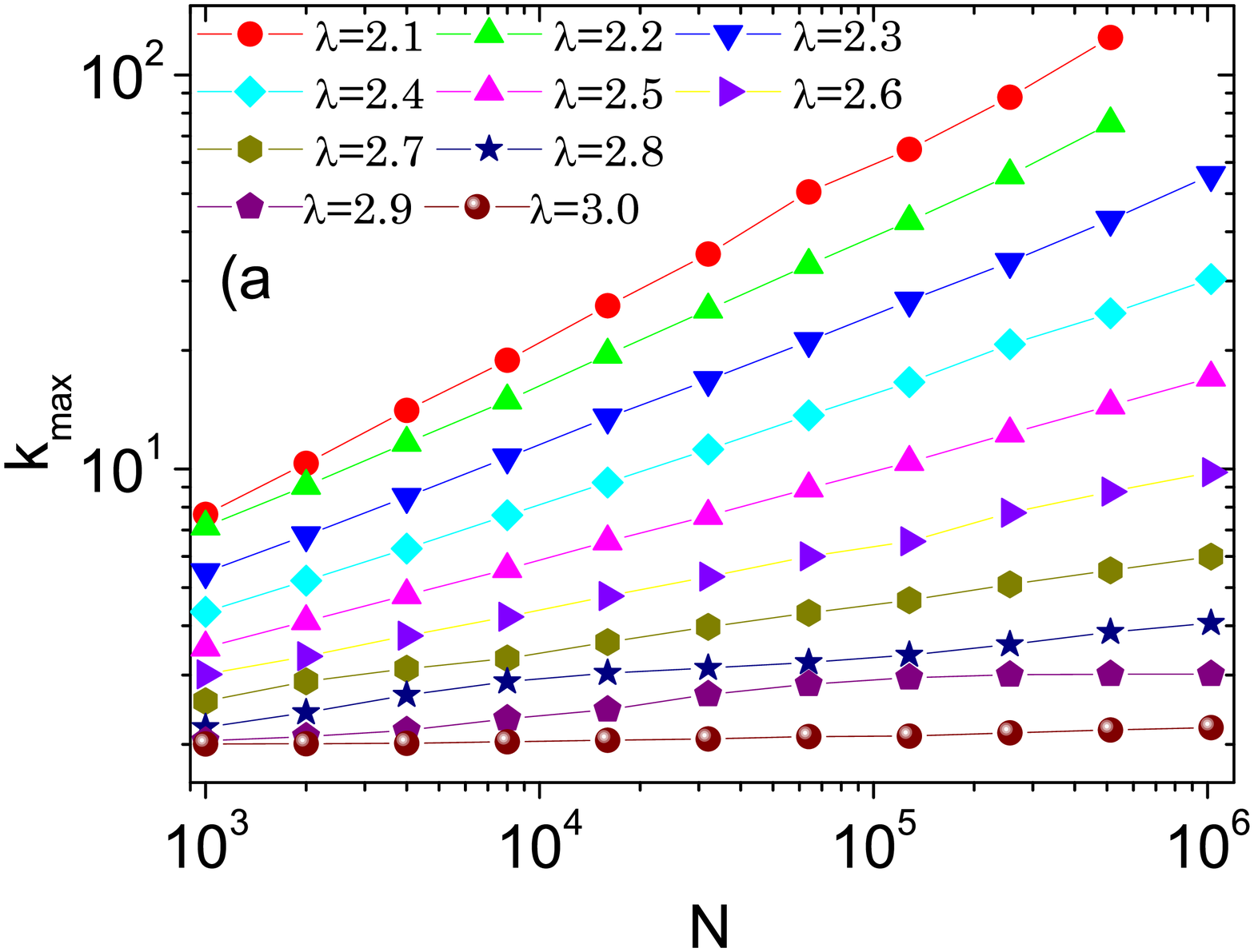}
\includegraphics[width=8.0 cm,height=6.9cm,angle=0]{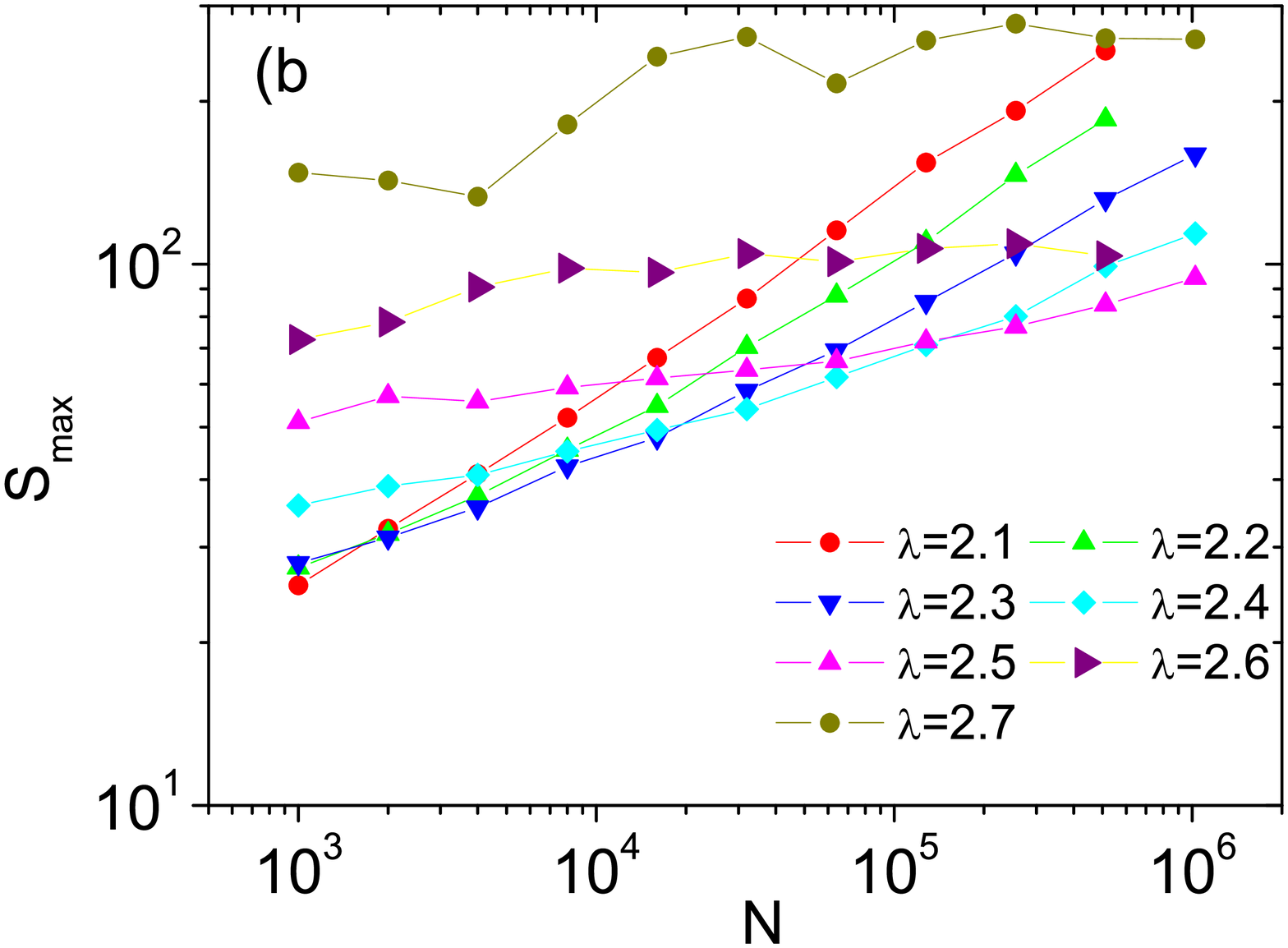}
\includegraphics[width=8.0 cm,height=6.9cm,angle=0]{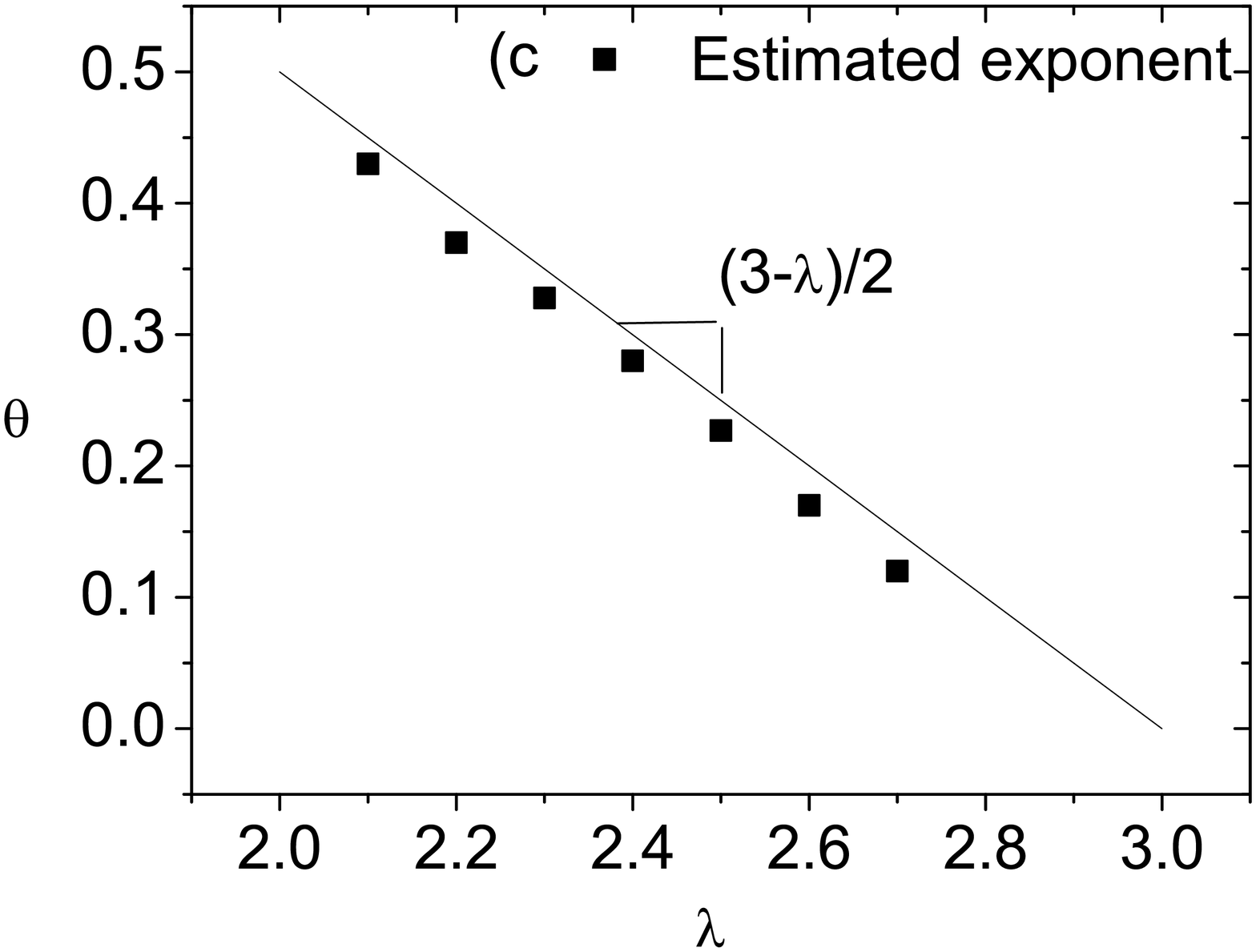}

\caption{(a) The total number of shells, $k_{\rm max}$, in SF models as
  a function of $N$. Note that $k_{\rm max} \propto N^{\theta} $. (b)
  The size of the last shell $S_{\rm max}$ in the SF model as a function
  of $N$. Each curve crosses over into a power law regime for $N \geq
  N_{c}(\lambda)$, where $N_{c}(\lambda)$ increases with $\lambda$. (c)
  The calculated exponent $\theta$ as a function of $\lambda$
  (symbols). Our calculated values of $\delta$ agree with the mean field
  theory result $\delta \approx 2/(3-\lambda)$ (solid line).}
\label{fig_kmax_nucl}
\end{figure}

One can estimate the total number of shells in a random SF network of size
$N$ as follows. Since every shell constitutes a fixed fraction of
$N$ it follows that $S_{k} \propto N k^{-\delta}$. The last shell
$k_{\rm max}$ needs to possess at least one node $S_{\rm max}\equiv
S_{k=k_{\rm max}} \sim 1$, which leads to
\begin{equation}
k_{\rm max} \leq N^{1/\delta}\label{exponent},
\end{equation}
Indeed, the total number of shells $k_{\rm max}$ seems to increase as a
power law with the network size $N$ [Fig.~\ref{fig_kmax_nucl}(a)]. The
smaller is $\lambda$ the faster is the growth of $k_{\rm max}$.  As seen
from Fig. \ref{fig_kmax_nucl}(c), the estimated exponents $1/\delta$
agree with those predicted by Eq.~(\ref{exponent}). Note that
Eq. (\ref{exponent}) together with
\begin{equation}
\delta = 2/(3-\lambda)
\end{equation}
is consistent with the fact that networks with $\lambda>3$ do not have a $k$-shell structure.

The dependence of $S_{\rm max}$ on $N$ can be regarded as a crossover
from the small $N$ regime with $N < N_{c}(\lambda)$ where there is no
nucleus to the power-law regime for $N > N_{c}(\lambda)$ where $S_{\rm
  max} \sim N^{\tau(\lambda)}$ ( Fig. \ref{fig_kmax_nucl}b). We relate
the observed crossover with the emergence of the nucleus and tendrils in
SF networks for $N>N_{c}(\lambda)$. The critical size of SF networks,
$N_{c}(\lambda)$, corresponding to the emergence of the nucleus and
tendrils in SF networks seems to increase as $\lambda$ increases. In the
$N < N_{c}$ regime the size of the last shell $S_{\rm max}$ increases
with $\lambda$, which can be explained by the fact that SF networks with
higher $\lambda$ have fewer shells. On the other hand, in the power-law
regime, the size of the last shell, $S_{\rm max}$, (which now becomes
the nucleus $S_{n}$) is smaller for larger values of $\lambda$.

\section{Evolution and structure of the LS and ICT industry networks}

\begin{figure} [!ht]
\includegraphics[width=8.0 cm,height=7.0cm,angle=0]{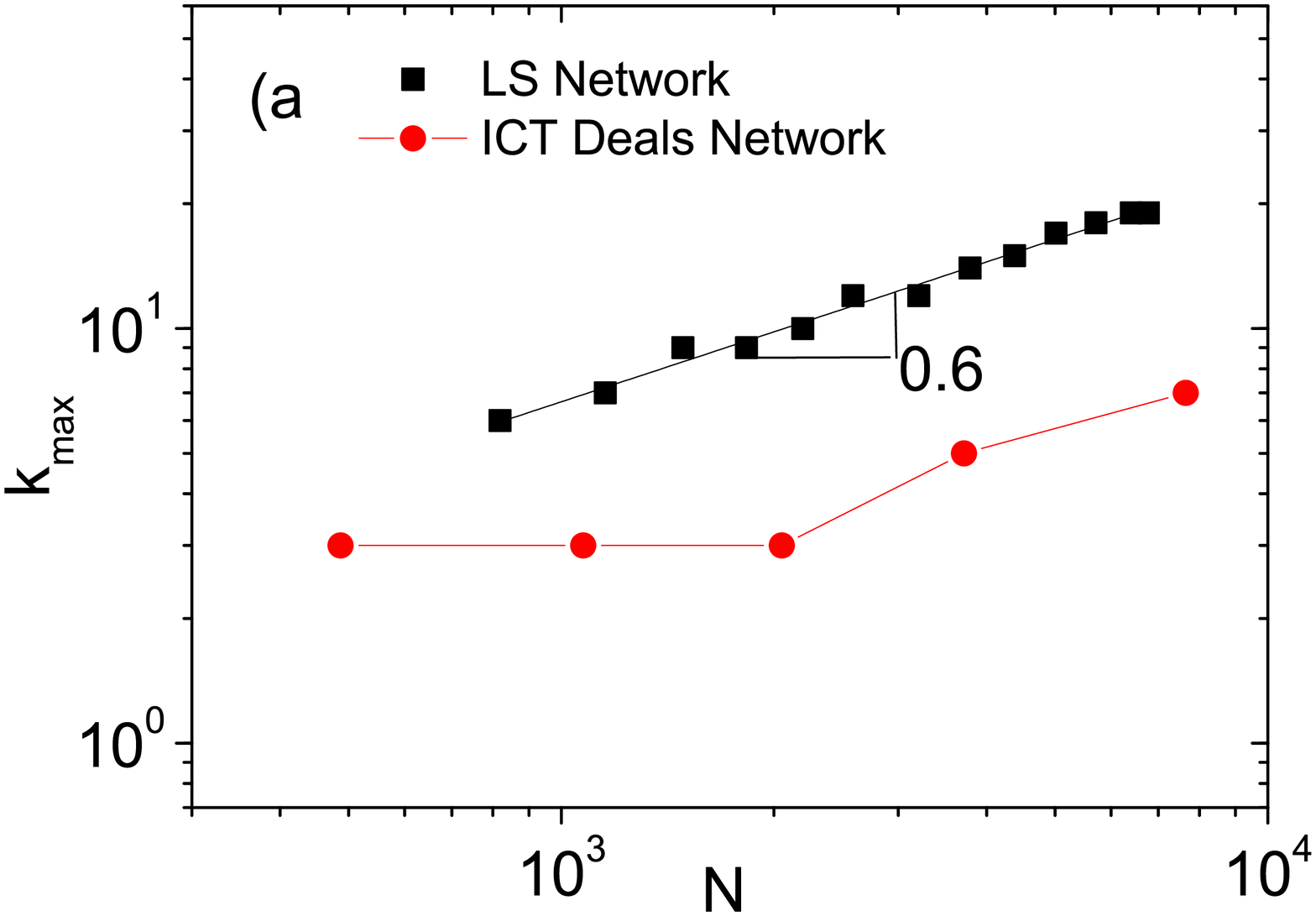}
\includegraphics[width=8.0 cm, height=6.9cm,angle=0]{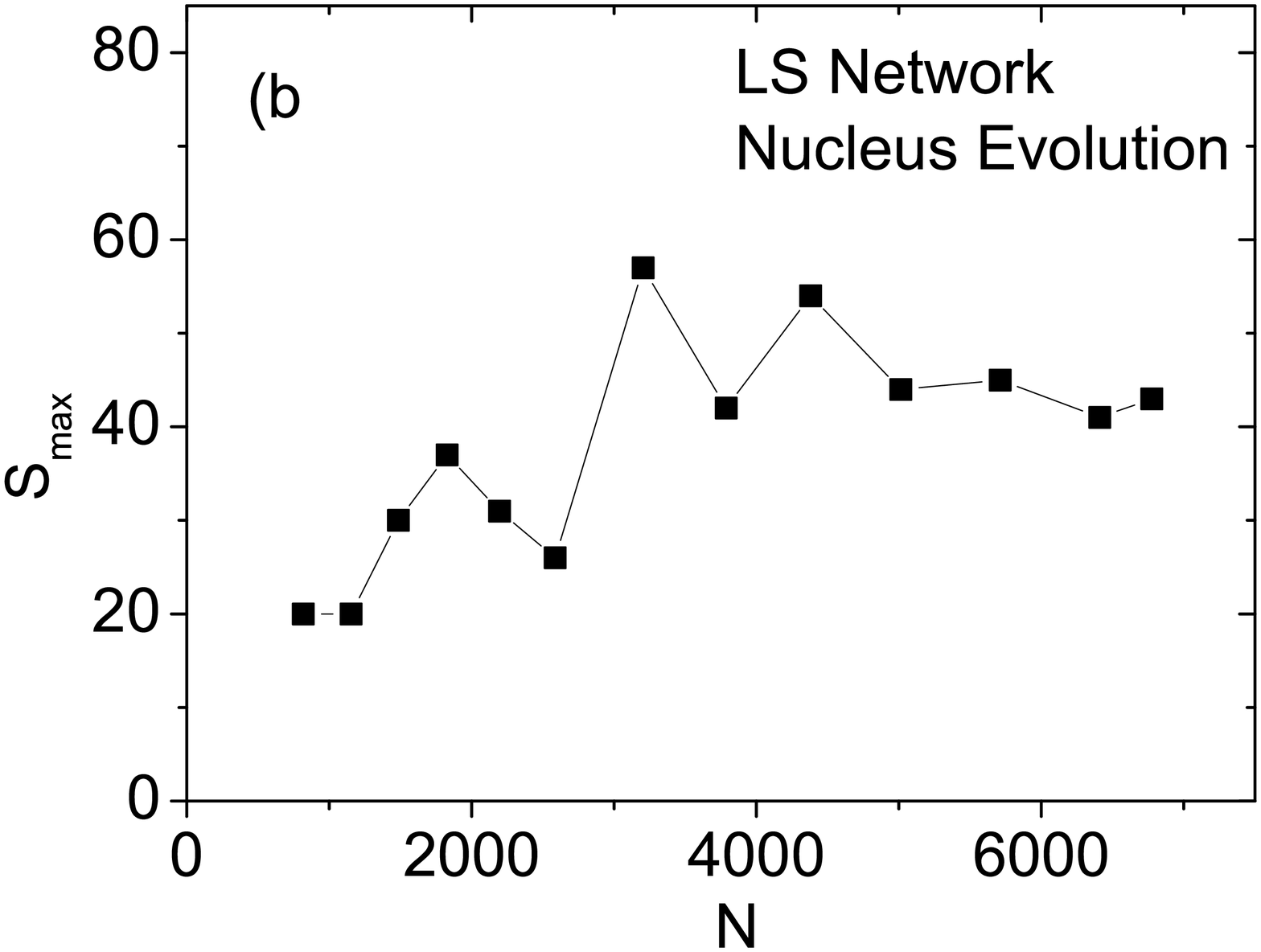}
\includegraphics[width=8.0 cm,height=7.0cm,angle=0]{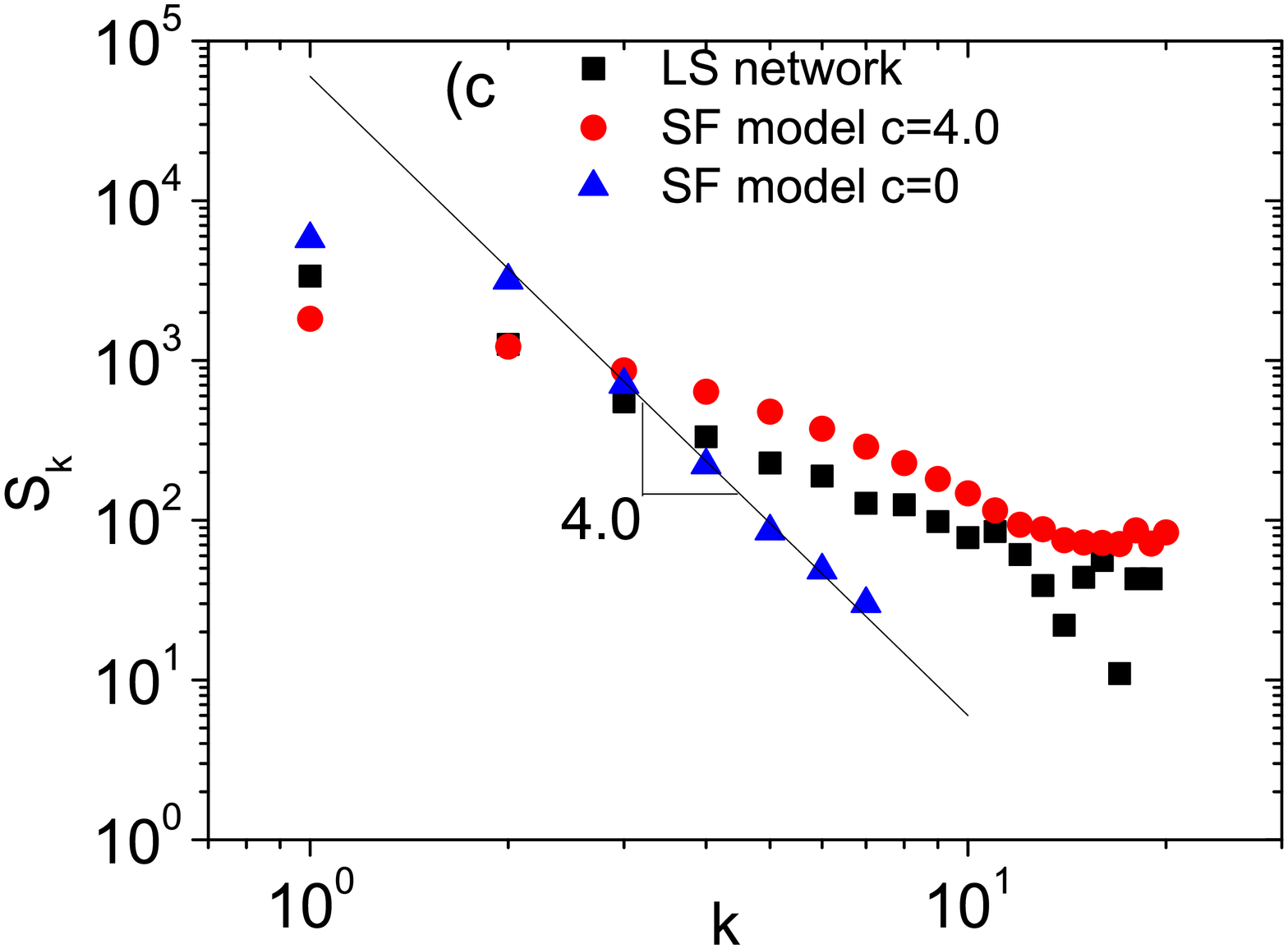}
\includegraphics[width=8.0 cm,height=7.0cm,angle=0]{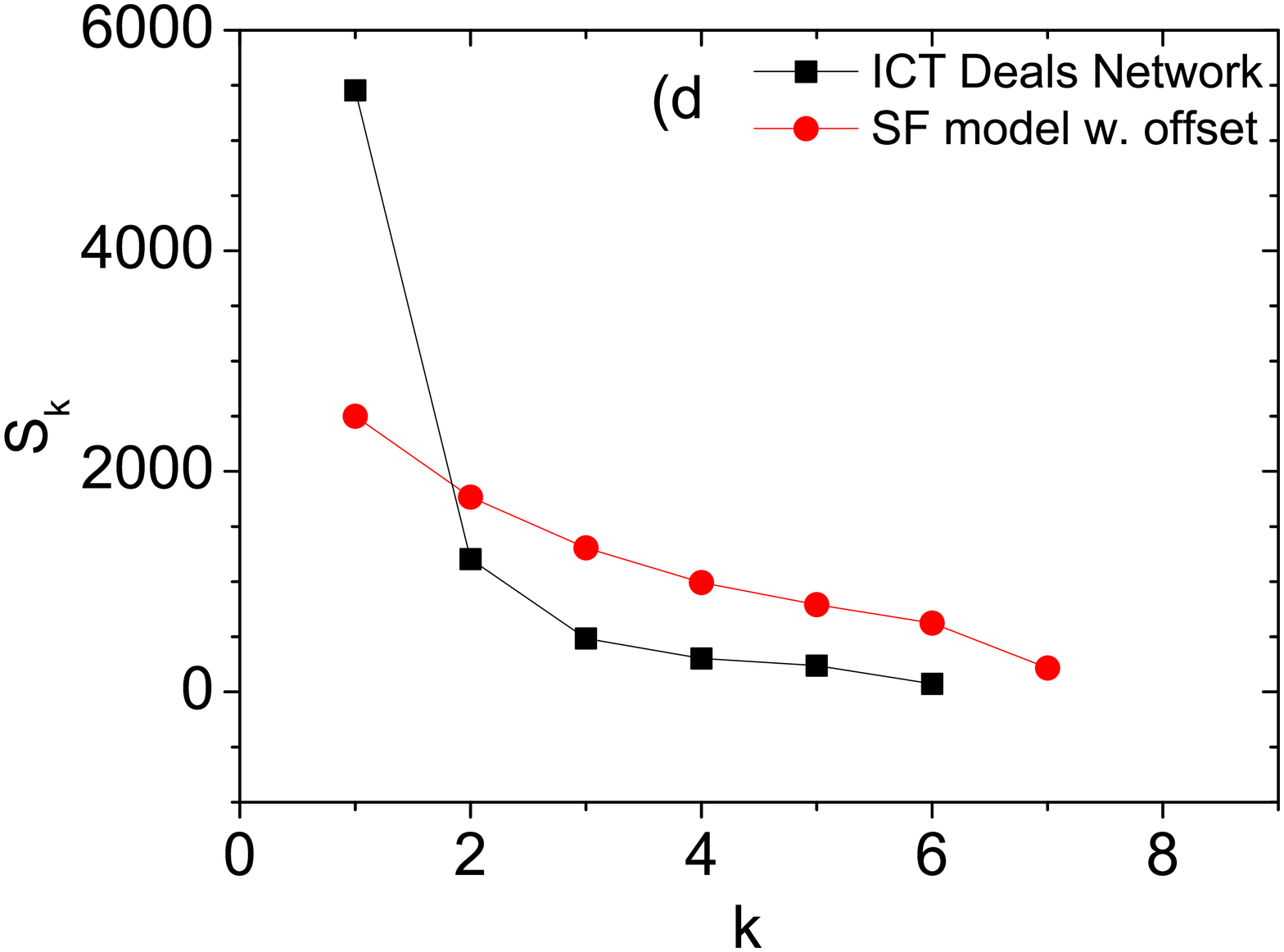}
\caption{(a) The largest shell index $k_{\rm max}$ of the LS and the ICT
  networks as a function of $N$ calculated for different years. The
  number of shells $k_{\rm max}$ of the LS network increases
  approximately as a power-law function of the network size $k_{\rm max}
  \sim N^{\theta}$, where $\theta \simeq 0.6$. (b) Size of the
  nucleus of the LS network, $S_{n}$, as a function of $N$. $S_{n}$
  exhibits fluctuations as LS network grows. Unlike in the analyzed SF
  models, $S_{\rm max}$ becomes stable for $N>5000$. (c) $S_{k}$ as a
  function of the $k$-shell index $k$ for the LS network (squares).
  Shell sizes $S_{k}$ as a function of $k$ decrease significantly slower
  than $S_{k} \sim k^{-4}$, which is expected for a random SF model with
  the same $\lambda$ and $c = 0$ (triangles). However, the offset
  introduction of $c=4.0$ in the SF degree distribution, $P(q)$, mimics
  the $k$-shell structure of the LS network (circles).  (d) $S_{k}$ as a
  function of $k$-shell index $k$ for the ICT network (squares). SF
  model with $\lambda = 3.4$ and $c=0$ does not possess a $k$-shell
  structure. However, the introduction of $c=6.0$ in the SF degree
  distribution yields similar $k$-shell structure (circles), but we
  attribute it only to finite size effect.}
\label{real_nws}
\end{figure}

We further analyze the $k$-shell structure of the LS and ICT industry networks.
The number of shells $k_{\rm max}$ in the LS industry grows as a
power-law function of its size,
\begin{equation}
k_{\rm max} \sim N^\theta,
\end{equation}
and reaches $k_{\rm max}=19$ in 2002. Our estimates yield $\theta
\approx 0.6$ (Fig. \ref{real_nws}a). We find that the number of shells
in ICT sector, $k_{\rm max}$, also grows and reaches $k_{\rm max}=6$ in
2000. Note that $S_{n}$ for the LS network exhibits fluctuations for
$N\in[2000,5000]$ and stabilizes for $N>5000$ (See
Fig. \ref{real_nws}b). As seen in Fig. \ref{real_nws}(c,d), the sizes of
shells $S_{k}$ decrease as a function of their index $k$ in both
networks.  As we notice in Section II, the observed shell sizes $S_{k}$ as
well as the number of shells $k_{\rm max}$, measured for the LS and ICT
networks, deviate from those obtained from SF models with
$c=0$. We expect for random SF models with $c=0$, $\lambda =2.5$ and
$\lambda = 3.4$ and similar sizes as LS and ICT to find $k_{\rm max}=5$
and $k_{\rm max} = 2$ respectively. Also, a SF network with
$\lambda=2.5$ is expected to have $S_{k}\sim k^{-4.}$ and $k_{\rm max}
\sim N^{0.25}$ which deviates from the observed shell
sizes [see Fig.~\ref{real_nws}(c)]. The observed differences can be
explained by taking into account the offset $c$ in SF degree
distribution of both networks. The adjustment of the degree distribution
of random SF models with $c=4.0$ and $c=6.0$ allows one to obtain
similar patterns $S_{k}$ which are in fair agreement with the industry
networks, as seen in Figs.~\ref{real_nws}(c) and \ref{real_nws}(d).

As seen above, the offset $c>0$ in the SF degree distribution plays a
crucial role in the formation of the $k$-shell structure of industry
networks. A possible reason for the emergence of $c>0$ in growing
networks is the combination of preferential attachment with random
attachment in network evolution \cite{pref}. The coexistence of
preferential attachment regime with random collaborative agreements was
suggested to take place in industry networks \cite{pharma}. The random
component is caused by the fact that sometimes firms choose exclusive
relationships and novelty, and do not prefer to make deals with hub
firms. Even though the coexistence of both preferential and random
regimes seems to be crucial in the formation of the industry networks,
it does not fully reproduce the $k$-shell structure of LS and ICT
industries. We believe, that a better understanding of the evolution of
the LS and ICT networks may be achieved by further improvements of the
modeling.

\section{Discussion and Summary}

We use the $k$-shell decomposition to analyze the structure of the LS and ICT sectors of industry.
We find that the firms in the LS industry can be naturally divided into three components: the nucleus, the tendrils and the bulk body.
The nucleus of the LS industry consists mostly of the market leader firms while the tendrils are typically comprised of small start-up firms
that preferentially sell their products to the nucleus firms. We show that the nucleus of the LS industry exhibits remarkable stability in time. In contrast, the ICT industry does not have a nucleus which can be explained by a high level of competition in the ICT sector.
We also analyzed the dependence of the $k$-shell structure of SF model networks
on $N$, $\lambda$ and $c$. We observed the formation of the nucleus and
the tendrils in SF networks only for $\lambda<3$.  The number of shells
$k_{\rm max}$ and the size of the nucleus $S_{n}$ are larger for SF
networks with $c>0$ compared to those with $c=0$. Our results can partly
explain the $k$-shell structure of LS and ICT industry networks. The
coexistence of preferential and random attachment leads to the
appearance of the offset $c>0$ in the SF degree distribution $P(q) \sim
(q+c)^{-\lambda}$ \cite{pref}. Thus, the appearance of $c>0$ in the degree
distribution of LS and ICT networks might be explained by the interplay
of random and preferential agreements among firms in the industries
\cite{pharma}.

\subsubsection*{Acknowledgments}

We thank ONR, European project DAPHNET,the Israel Complexity Center,
Merck Foundation (EPRIS project) and Israel Science Foundation for
financial support. We thank L. Braunstein, S. Carmi and L. K. Gallos for valuable discussions.

\newpage
\begin{figure}[!h]
\caption{Illustration of the $k$-shell decomposition method. (a)
  Original network. Nodes are marked by corresponding $k$-shell
  indices. Note that the $k$-shell index does not coincide with the node
  degree. (b) The 3-crust (left) and the 3-core (right) of the original
  network.} \label{drawing}
\end{figure}


\end{document}